# EM-based Fast Uncertainty Quantification for Bayesian Multi-setup Operational Modal Analysis

Wei Zhu[a,b] , Binbin Li[a,b*] and Zuo Zhu[c]

[a] State Key Laboratory of Biobased Transportation Fuel Technology, ZJU-UIUC Institute, Zhejiang University, Haining, 314400, China.

[b]College of Civil Engineering and Architecture, Zhejiang University Hangzhou, 310058, China.

[c]Vibration Engineering Section, College of Engineering, Mathematics and Physical Sciences, University of Exeter, Exeter, UK

**Abstract:** The current Bayesian FFT algorithm relies on direct differentiation to obtain the posterior covariance matrix (PCM), which is time-consuming, memory-intensive, and hard to code, especially for the multi-setup operational modal analysis (OMA). Aiming at accelerating the uncertainty quantification in multi-setup OMA, an expectation-maximization (EM)-based algorithm is proposed by reformulating the Hessian matrix of the negative log-likelihood function (NLLF) as a sum of simplified components corresponding to the complete-data NLLF. Matrix calculus is employed to derive these components in a compact manner, resulting in expressions similar to those in the single-setup case. This similarity allows for the reuse of existing Bayesian single-setup OMA codes, simplifying implementation. The singularity caused by mode shape norm constraints is addressed through null space projection, eliminating potential numerical errors from the conventional pseudoinverse operation. A sparse assembly strategy is further adopted, avoiding unnecessary calculations and storage of predominant zero elements in the Hessian matrix. The proposed method is then validated through a comprehensive parametric study and applied to a multi-setup OMA of a high-rise building. Results demonstrate that the proposed method efficiently calculates the PCM within seconds, even for cases with hundreds of parameters. This represents an efficiency improvement of at least one order of magnitude over the state-of-the-art method. Such performance paves the way for a real-time modal identification of large-scale structures, including those with closely-spaced modes.

[*] Corresponding author: bbl@zju.edu.cn

# 1 Introduction

Operational Modal Analysis (OMA) is a widely adopted technique for identifying dynamic properties of structures, such as natural frequencies, damping ratios, and mode shapes, under operational conditions [1-3]. It is ideal for the continuous monitoring of infrastructure where artificial excitation is either impractical or undesirable [4, 5]. OMA has found broad applications in areas such as structural vibration control [6, 7], damage detection [8-10], and model updating [11-13], underscoring its importance across various engineering disciplines. Usually, OMA supposes that all degrees of freedom (DoFs) are measured synchronously in a single setup. Several methods have been developed to extract modal parameters and quantify uncertainty under this assumption.

From the perspective of the frequentist (classical) statistics, the two widely used and representative OMA methods are stochastic subspace identification (SSI), and frequency domain decomposition (FDD). The former was proposed by Overschee and Moor [14], operating in the time domain, while the latter was developed by Brincker et al. [15], working in the frequency domain. Both methods construct data-driven estimators as proxies for the modal parameters and interpret uncertainty through the variability in repeated experiments. Comprehensive uncertainty quantification methods for the SSI were established by Reynders et al. [16, 17]. However, a comparable uncertainty quantification framework for FDD has yet to be developed.

Alternatively, the Bayesian (non-classical) approach incorporates uncertainty probabilistically, considering modal parameters as random variables with probability density functions (PDF) informed by data and modeling assumptions. Yuen and Katafygiotis [18] first introduced this approach in the time domain, which was later extended to the frequency domain [19, 20]. Although early implementation was computationally expensive, the development of the fast Bayesian fast Fourier transform (FFT) algorithm by Au [21] significantly enhanced its practicality. Au [22] further proposed a more general version, allowing for efficient practical implementation and robust uncertainty quantification. Building on these advances, Li and Au [23], introduced the expectation-maximization (EM) algorithm, which laid the foundation for further improvements by Zhu et al. [24]. Additionally, the release of public source code by Zhu and Li [25] has facilitated the broader adoption and practical application of this technique. However, single-setup OMA may still not be sufficient when a detailed mode shape with more DoFs is required than can be measured at once.

To address this limitation, the multi-setup test is employed to distribute sensors across different setups. Traditional multi-setup OMA employs two primary strategies to handle the distributed data: pre-identification and post-identification. In the former, data from different setups are combined into a single dataset for analysis, assuming that modal parameters remain invariant across all setups. A multi-setup version of SSI falls into this category [26], and detailed uncertainty analyses for this method can be found in [27]. Nevertheless, such invariance assumption proves to be restrictive, especially when long stationary processes are required, leading to potential modeling errors [28].

In contrast, the post-identification approach processes data from each setup individually, deriving ‘local’ mode shapes for each setup and then combining them to form the “global” mode shapes in a least-squares manner. This heuristic method, however, heavily relies on the overall data quality, making it less robust when the signal-to-noise ratio (SNR) is low in some setups [2]. To improve this situation, Au [29] proposed a global least-squares method, transforming the problem into a constrained optimization task. This approach allows for a more comprehensive weighting of different setups, improving the robustness of the solution. While it often yields physically reasonable results, the practical implementation is restricted due to its inability to quantify the uncertainty of the identified mode shapes.

To overcome the limitations of conventional approaches, the multi-setup Bayesian FFT method was introduced. Unlike pre- or post-identification methods, it does not assume consistent modal parameters across setups or depend solely on local mode shapes for global estimates. Instead, it treats the global mode shape as one of the model parameters within a Bayesian framework, effectively linking all parameters and datasets through a well-constructed negative log-likelihood function (NLLF). However, the expanded parameter space and large datasets introduce additional challenges. Initially, Au [30] proposed this method to handle the case with well-separated modes. Zhang et al. [31] subsequently introduced an uncertainty quantification framework to enhance its completeness. To better understand and manage uncertainty, Xie et al. [32] then developed an "uncertainty law" for multi-setup OMA with well-separated modes, analytically illustrating how identification uncertainty is influenced by test configuration. Later, Zhu et al. [33] expanded the applicability of the multi-setup Bayesian FFT method to scenarios involving closely-spaced modes by an EM algorithm. Although they [34] also developed associated uncertainty quantification techniques, the approach still struggles with poor computational efficiency and high memory demands due to the large number of parameters and datasets involved.

To address these challenges, this paper presents an EM-based fast method for posterior covariance matrix (PCM) computation, aiming at efficiently quantifying identification uncertainty in multi-setup Bayesian FFT analysis. The proposed method first employs a null-space projection to handle singularities caused by mode shape norm constraints, thereby eliminating potential numerical instability. A sparse assembly strategy is then applied to optimize memory usage and computational loads by avoiding unnecessary storage and calculation of zero elements in the Hessian matrix. To further reduce computational complexity, we introduce a latent variable to simplify the structure of the NLLF. Instead of calculating second-order derivatives directly, Louis' identity is adopted to derive the necessary components. With enhanced computational efficiency and robust uncertainty quantification, this method paves the way for real-time modal analysis of large-scale structures, especially in scenarios involving closely-spaced modes.

# 2 Review on multi-setup Bayesian FFT method

This section reviews the Bayesian FFT method incorporating multiple setups to provide readers with the essential background of the proposed method. It begins by introducing the Bayesian modeling process for datasets obtained from multi-setup vibration tests, covering aspects such as FFT data modeling, Bayes' theorem application, and Laplace posterior approximation. Following this, the Bayesian model is reformulated as a latent variable model, and the fundamental concept of implementing the EM algorithm is introduced.

## 2.1 Probabilistic model

In multi-setup ambient vibration tests, some reference DoFs, are measured repeatedly across different setups. Meanwhile, sensors are moved to capture other positions, referred to as rover DoFs, to ensure comprehensive coverage of the structure. Assuming an ambient vibration test is conducted in $n_s$ setups, with $\left\{\hat{\boldsymbol{y}}_j^{(r)} \in \mathbb{R}^{n_r}: j = 0, 1, \cdots, N_r - 1\right\}$ representing the collected time history data. In this paper, unless otherwise specified, quantities with superscript $[\cdot]^{(r)}$ or subscript $[\cdot]_r$ indicate evaluation in Setup $r$ ($r = 1, 2, \cdots, n_s$). Here, $N_r$ is the number of samples per data channel, while $n_r$ is the number of data channels. By applying a two-sided scaled FFT at $\mathrm{f}_k^{(r)} = k/N_r \Delta t_r$ (Hz), the frequency-domain representation is given by

$$\hat{\boldsymbol{\mathcal{F}}}_k^{(r)} = \sqrt{\frac{\Delta t_r}{N_r}} \sum_{j=0}^{N_r-1} \hat{\boldsymbol{y}}_j^{(r)} e^{-\mathbf{i}2\pi jk/N_r} \tag{1}$$

where $\Delta t_r$ is the sampling interval; $\mathbf{i} = \sqrt{-1}$ is the imaginary unit; $k = 0, 1, \cdots, \mathrm{int}[N_r/2]$ ($\mathrm{int}[\cdot]$ denotes the integer part).

Focusing on a narrow band $\left[\mathrm{f}_\mathrm{l}^{(r)}, \mathrm{f}_\mathrm{u}^{(r)}\right]$ around the resonance peaks, the Bayesian FFT method models the FFT data $\boldsymbol{\mathcal{D}}^{(r)} = \left\{\hat{\boldsymbol{\mathcal{F}}}_k^{(r)} \in \mathbb{C}^{n_r \times 1}\right\}$ within this band as the sum of two parts,

$$\hat{\boldsymbol{\mathcal{F}}}_k^{(r)} = \boldsymbol{\Phi}_r \mathbf{h}_k^{(r)} \boldsymbol{p}_k^{(r)} + \boldsymbol{\varepsilon}_k^{(r)} \tag{2}$$

where the first part is the frequency response of an assumed linear system, validated by the modal superposition theorem [35]; The second term $\boldsymbol{\varepsilon}_k^{(r)} \in \mathbb{C}^{m \times 1}$ accounts for prediction error due to data noise and modeling error. Typically, $m$, the number of modes dominating the response in this band, is no more than 3. Here, $\boldsymbol{\Phi}_r \in \mathbb{R}^{n_r \times m}$ is the 'local' mode shape confined to the partially measured DoFs in Setup $r$; $\boldsymbol{p}_k^{(r)} \in \mathbb{C}^{m \times 1}$ is the modal force normalized by modal mass; $\mathbf{h}_k^{(r)} = \mathrm{diag}\left(\left[h_{1k}^{(r)}, h_{2k}^{(r)}, \cdots, h_{mk}^{(r)}\right]\right) \in \mathbb{C}^{m \times m}$ is a diagonal matrix where each diagonal term represents the frequency response function

$$h_{1k}^{(r)} = \frac{\left(\mathbf{i}2\pi \mathrm{f}_k^{(r)}\right)^{-q}}{\left(1 - \beta_{ik}^{(r)2}\right) - 2\mathbf{i}\zeta_i^{(r)}\beta_{ik}^{(r)}}; \quad \beta_{ik}^{(r)} = \frac{f_i^{(r)}}{\mathrm{f}_k^{(r)}} \tag{3}$$

where $\mathrm{diag}(\cdot)$ transforms a vector into a diagonal matrix; $f_i^{(r)}$ and $\zeta_i^{(r)}$ denote the natural frequency and damping ratio of the $i$-th mode ($i = 1, 2, \cdots, m$), respectively; $q = 0, 1, 2$ corresponds to acceleration, velocity, and displacement data, respectively.

Since the main target for the multi-setup test is to obtain the global mode shape $\boldsymbol{\Phi} = [\boldsymbol{\varphi}_1, \boldsymbol{\varphi}_2, \cdots, \boldsymbol{\varphi}_m] \in \mathbb{R}^{n \times m}$ covering all measured DoFs $n$, a selection matrix $\mathbf{C}_r$ is introduced so that

$$\mathbf{\Phi}_r = \mathbf{C}_r\mathbf{\Phi} \tag{4}$$

where $\mathbf{C}_r \in \mathbb{R}^{n_r \times n}$ is a permutation matrix determined by the test configuration. If the $j$-th DoF in $\mathbf{\Phi}$ is measured by the $i$-th data channel in Setup $r$, the $(i, j)$-entry of $\mathbf{C}_r$ is one, otherwise zero.

In the Bayesian FFT method, $\boldsymbol{p}_k^{(r)}$ and $\boldsymbol{\varepsilon}_k^{(r)}$ are assumed to be zero-mean independent stochastic stationary processes with constant power spectral density (PSD) $\mathbf{S}^{(r)} \in \mathbb{C}^{m \times m}$ and $S_{\mathrm{e}}^{(r)}\mathbf{I}_{n_r} \in \mathbb{R}^{n_r \times n_r}$ within the selected band. For large $N_r$ ($\gg 1$), $\hat{\boldsymbol{\mathcal{F}}}_k^{(r)}$ becomes asymptotically independent across different frequencies and follows a zero-mean complex Gaussian distribution. The covariance matrix is expressed as $\mathbf{E}_k^{(r)} = \mathbf{\Phi}_r\mathbf{H}_k^{(r)}\mathbf{\Phi}_r^{\mathrm{T}} + S_{\mathrm{e}}^{(r)}\mathbf{I}_{n_r} \in \mathbb{C}^{n_r \times n_r}$, where $\mathbf{H}_k^{(r)} = \mathbf{h}_k^{(r)}\mathbf{S}^{(r)}\mathbf{h}_k^{(r)*}$; $[\cdot]^{\mathrm{T}}$ and $[\cdot]^*$ denote matrix transpose and conjugate, respectively. Without loss of generality, let $\boldsymbol{\theta} = \left[\boldsymbol{x}^{(1)\mathrm{T}}, \boldsymbol{x}^{(2)\mathrm{T}}, \cdots, \boldsymbol{x}^{(n_s)\mathrm{T}}, \mathrm{vec}^{\mathrm{T}}(\mathbf{\Phi})\right]^{\mathrm{T}}$ denote the real-valued vector of modal parameters to be identified, where $\mathrm{vec}(\cdot)$ denotes the column-wise stacking of matrix elements (see Appendix A). Here, $\boldsymbol{x}^{(r)} = \left[\boldsymbol{f}^{(r)\mathrm{T}}, \boldsymbol{\zeta}^{(r)\mathrm{T}}, \mathrm{vec}^{\mathrm{T}}\left[\mathbf{S}^{(r)}\right], S_{\mathrm{e}}^{(r)}\right]^{\mathrm{T}} \in \mathbb{R}^{n_{\theta_r} \times 1}$ defines the set of independent parameters for Setup $r$ and $n_{\theta_r} = (m+1)^2$ with $m$ variables for $\boldsymbol{f}^{(r)}$, $m$ for $\boldsymbol{\zeta}^{(r)}$, $m^2$ for $\mathbf{S}^{(r)}$ and, 1 for $S_e^{(r)}$, respectively. Consequently, the total number of parameters is $n_\theta = n_s(m+1)^2 + mn$, where $n$ represents the total measured DoFs across all setups. Assuming FFT data are independent among setups, the NLLF $L(\boldsymbol{\theta}) = -\ln p\left(\left\{\boldsymbol{\mathcal{D}}^{(1)}, \boldsymbol{\mathcal{D}}^{(2)}, \cdots, \boldsymbol{\mathcal{D}}^{(n_s)}\right\}\middle|\boldsymbol{\theta}\right) = \sum_{r=1}^{n_s}\sum_k\left[L_k^{(r)}\right]$ with

$$L_k^{(r)} = n_r \ln \pi + \hat{\boldsymbol{\mathcal{F}}}_k^{(r)\mathrm{H}}\left[\mathbf{E}_k^{(r)}\right]^{-1}\hat{\boldsymbol{\mathcal{F}}}_k^{(r)} + \ln\left|\mathbf{E}_k^{(r)}\right| \tag{5}$$

where $\ln[\cdot]$ is the natural logarithm; $[\cdot]^{\mathrm{H}}$ denotes the Hermitian transpose of a complex-valued matrix; $[\cdot]^{-1}$ denotes the inverse; $|\cdot|$ denotes the determinant.

Adopting a uniform prior and a Laplace approximation, the posterior PDF (given data) can be approximated by a multivariate normal distribution with the mean and covariance matrix given by

$$\hat{\boldsymbol{\theta}} = \arg\min_{\boldsymbol{\theta}} L(\boldsymbol{\theta}) \in \mathbb{R}^{n_\theta \times 1} \tag{6}$$

$$\hat{\mathbf{C}} = \left[\nabla^2 \hat{L}(\boldsymbol{\theta})\right]^{-1} \in \mathbb{R}^{n_\theta \times n_\theta} \tag{7}$$

where $\arg\min_{\boldsymbol{\theta}}[\cdot]$ represents the value that minimizes $L(\boldsymbol{\theta})$ and $\nabla^2[\cdot]$ denotes the Hessian matrix of the respective function. The posterior mean here is also known as the most probable value (MPV) of modal parameters. The $\nabla^2\hat{L}(\boldsymbol{\theta})$ indicates that the Hessian matrix is evaluated at $\boldsymbol{\theta} = \hat{\boldsymbol{\theta}}$. Note that this Bayesian model is now unidentifiable because the mode shape can be arbitrarily normalized. Euclidean norm constraints are therefore imposed on the mode shape, i.e., $\mathbf{G}(\boldsymbol{\theta}) = [g_1(\boldsymbol{\theta}), g_2(\boldsymbol{\theta}), \cdots, g_m(\boldsymbol{\theta})]^{\mathrm{T}}$ with

$$g_i(\boldsymbol{\theta}) = 0.5\left(\boldsymbol{\varphi}_i\boldsymbol{\varphi}_i^{\mathrm{T}} - 1\right) = 0 \quad (i = 1, 2, \cdots, m) \tag{8}$$

This makes the MPV computation a constrained optimization problem. Complicated by the coupling of parameters, direct optimization [30] becomes inefficient. In this respect, the EM algorithm offers an efficient alternative, as detailed below.

## 2.2 EM algorithm

The EM algorithm [36] is a powerful statistical technique for parameter estimation in probabilistic models. This algorithm is efficient in scenarios where data are incomplete or contain latent variables [37]. Due to the unmeasured nature of the excitation in OMA, quantities related to it can be modeled as latent variables. Following the procedures proposed by Li and Zhu [23, 33], the modal responses $\boldsymbol{\eta}_k^{(r)} = \mathbf{h}_k^{(r)}\boldsymbol{p}_k^{(r)}$ among setups are considered as a set of latent variables. In this way, the NLLF transforms into its 'complete-data' counterpart, which is given by $\mathcal{L}(\boldsymbol{\theta}) = \sum_{r=1}^{n_s}\sum_k \left[\mathcal{L}_k^{(r)}(\boldsymbol{\theta})\right]$ with

$$\begin{aligned}\mathcal{L}_k^{(r)}(\boldsymbol{\theta}) = (m + n_r)\ln\pi + n_r\ln S_{\mathrm{e}}^{(r)} + 1/S_{\mathrm{e}}^{(r)}\left(\hat{\boldsymbol{\mathcal{F}}}_k^{(r)} - \boldsymbol{\Phi}_r\boldsymbol{\eta}_k^{(r)}\right)^{\mathrm{H}}\left(\hat{\boldsymbol{\mathcal{F}}}_k^{(r)} - \boldsymbol{\Phi}_r\boldsymbol{\eta}_k^{(r)}\right) \\ + \boldsymbol{\eta}_k^{(r)\mathrm{H}}\left[\mathbf{H}_k^{(r)}\right]^{-1}\boldsymbol{\eta}_k^{(r)} + \ln\left|\mathbf{H}_k^{(r)}\right|\end{aligned} \tag{9}$$

The EM algorithm operates in two main steps to solve this latent variable model: the Expectation step (E-step) and the Maximization step (M-step). In the E-step, the algorithm estimates the expected values of the complete-data NLLF for the distribution $p\left(\boldsymbol{\eta}_k^{(r)}\middle|\hat{\boldsymbol{\mathcal{F}}}_k^{(r)}, \boldsymbol{\theta}\right)$, referred to as the Q-function $Q_k^{(r)}(\boldsymbol{\theta}) = \mathrm{E}_{\boldsymbol{\theta}}\left[\mathcal{L}_k^{(r)}(\boldsymbol{\theta})\right]$. Here, E[·] denotes the expectation operator. Because the conditional distribution of a complex Gaussian distribution remains Gaussian [38], $p\left(\boldsymbol{\eta}_k^{(r)}\middle|\hat{\boldsymbol{\mathcal{F}}}_k^{(r)}, \boldsymbol{\theta}\right)$ follows a complex Gaussian distribution with the mean and covariance matrix given by

$$\boldsymbol{w}_k^{(r)} = \mathrm{E}\left[\boldsymbol{\eta}_k^{(r)}\right] = \left[\mathbf{P}_k^{(r)}\right]^{-1}\boldsymbol{\Phi}_r^{\mathrm{T}}\hat{\boldsymbol{\mathcal{F}}}_k^{(r)} \in \mathbb{C}^{m\times 1} \tag{10}$$

$$\boldsymbol{\Sigma}_k^{(r)} = \mathrm{E}\left[\left(\boldsymbol{\eta}_k^{(r)} - \boldsymbol{w}_k^{(r)}\right)\left(\boldsymbol{\eta}_k^{(r)} - \boldsymbol{w}_k^{(r)}\right)^{\mathrm{H}}\right] = S_{\mathrm{e}}^{(r)}\left[\mathbf{P}_k^{(r)}\right]^{-1} \in \mathbb{C}^{m\times m} \tag{11}$$

where $\mathbf{P}_k^{(r)} = S_{\mathrm{e}}^{(r)}\left[\mathbf{H}_k^{(r)}\right]^{-1} + \boldsymbol{\Phi}_r^{\mathrm{T}}\boldsymbol{\Phi}_r$. Combining Eqns. (9)~(11), the Q-function is given by

$$\begin{aligned}Q_k^{(r)}(\boldsymbol{\theta}) = (m + n_r)\ln\pi + n_r\ln S_{\mathrm{e}}^{(r)} + 1/S_{\mathrm{e}}^{(r)}\hat{\boldsymbol{\mathcal{F}}}_k^{(r)\mathrm{H}}\hat{\boldsymbol{\mathcal{F}}}_k^{(r)} + \mathrm{tr}\left[\mathrm{Re}\left(\left[\mathbf{H}_k^{(r)}\right]^{-1}\mathbf{W}_k^{(r)}\right)\right] \\ + \ln\left|\mathbf{H}_k^{(r)}\right| - 2/S_{\mathrm{e}}^{(r)}\,\mathrm{tr}\left[\boldsymbol{\Phi}_r\mathrm{Re}\left(\boldsymbol{w}_k^{(r)}\hat{\boldsymbol{\mathcal{F}}}_k^{(r)\mathrm{H}}\right)\right] + 1/S_{\mathrm{e}}^{(r)}\,\mathrm{tr}\left[\boldsymbol{\Phi}_r\mathrm{Re}\left(\mathbf{W}_k^{(r)}\right)\boldsymbol{\Phi}_r^{\mathrm{T}}\right]\end{aligned} \tag{12}$$

where we have defined $\mathbf{W}_k^{(r)} = \hat{\boldsymbol{w}}_k^{(r)}\hat{\boldsymbol{w}}_k^{(r)\mathrm{H}} + \hat{\boldsymbol{\Sigma}}_k^{(r)}$; $\mathrm{tr}(\cdot)$ refers to the trace of a matrix, i.e., the sum of its diagonal entries; $\mathrm{Re}(\cdot)$ denotes taking the real part of a matrix.

In the M-step, the algorithm updates the parameters to minimize the Q-function based on these estimates. Here, the modal parameters are found to be partially decoupled in the complete-data NLLF and the Q-function. This desirable property allows some parameters, such as the global mode shape $\mathbf{\Phi}$, to be updated analytically, reducing the complexity of the overall process.

Given an initial point, this process repeats until the convergence criterion is satisfied. The EM algorithm is efficient during early iterations but shows little improvement as it approaches the final solution. As a result, the identification accuracy is not acceptable solely depending on it. Successful implementation of its variants such as parabolic extrapolation EM (P-EM) [23] makes the convergence result more robust and reliable. However, P-EM is not available as the final MPV, as the imposed constraints are not included in the optimization process. Note that the constraints only serve to enforce the unity Euclidean norm on the mode shape, and this can be addressed in a post-processing step. Specifically, let $\hat{\mathbf{\Phi}}'$ and $\left\{\hat{\mathbf{S}}^{(r)\prime}\right\}_{r=1}^{n_s}$ be the results obtained from the P-EM algorithm. Then define the column norms of $\hat{\mathbf{\Phi}}'$ as diagonal matrix $\mathbf{D} \in \mathbb{R}^{m \times m}$ and employ the renormalization

$$\hat{\mathbf{\Phi}} = \hat{\mathbf{\Phi}}'\mathbf{D}^{-1}; \left\{\hat{\mathbf{S}}^{(r)}\right\}_{r=1}^{n_s} = \mathbf{D}\left\{\hat{\mathbf{S}}^{(r)\prime}\right\}_{r=1}^{n_s}\mathbf{D} \tag{13}$$

In this way, the unity column norm of the mode shape matrix is ensured while preserving the NLLF value. This can be verified through the covariance matrix $\mathbf{E}_k^{(r)}$ in Eqn. (5).

Based on the MPV result, the next goal of the Bayesian FFT method is to calculate the PCM to quantify the identification uncertainty. This task is challenging due to the constraints shown in Eqn. (8). While handing the NLLF directly is theoretically feasible, it proves to be complex in terms of derivation and coding. A more efficient idea is to make use of the decoupled nature of the complete-data NLLF, as utilized in the MPV computation. This is introduced in the next section.

# 3 EM-based fast PCM computation

In this section, we propose a new method for calculating the PCM, structured into three key steps. First, the Hessian matrix of the NLLF is efficiently inverted, addressing the singularities caused by norm constraints. A sparse assembly method for the Hessian matrix is further adopted, improving both computational efficiency and memory usage. Finally, the essential parts of the Hessian matrix are computed by decomposing the second-order derivatives of the NLLF with respect to (w. r. t.) the modal parameters into lower-dimensional terms. These components are derived using results from the single-setup case, streamlining the overall process.

## 3.1 Robust norm constraints handing

As Au and Xie [39] proposed, the PCM of $\boldsymbol{\theta}$ under constraints can be indirectly calculated through a set of free (unconstrained) variables $\boldsymbol{u}$, as

$$\hat{\mathbf{C}} = \nabla\hat{\boldsymbol{v}}_{\mathrm{c}}(\boldsymbol{u})\left[\nabla^2\hat{L}_{\mathrm{c}}(\boldsymbol{u})\right]^{\dagger}\nabla^{\mathrm{T}}\hat{\boldsymbol{v}}_{\mathrm{c}}(\boldsymbol{u}) \tag{14}$$

where $\nabla[\cdot]$ denotes the gradient of the quantity; the mapping function is defined as $\boldsymbol{\theta} = \boldsymbol{v}_{\mathrm{c}}(\boldsymbol{u})$ so that the constraints in Eqn. (8) can be automatically satisfied, i.e., $g_i(\boldsymbol{v}_{\mathrm{c}}(\boldsymbol{u})) = 0$ for any $\boldsymbol{u}$.The function $L_{\mathrm{c}}(\boldsymbol{u}) = L(\boldsymbol{v}_{\mathrm{c}}(\boldsymbol{u}))$ is a composite function by rewriting the NLLF in terms of $\boldsymbol{u}$. Physically, the transformation from the unconstrained mode shape to the constrained one reflects a scaling of the mode shape. $L_{\mathrm{c}}(\boldsymbol{u})$ is proved to be invariant to the choice of $\boldsymbol{u}$ and $\boldsymbol{v}_{\mathrm{c}}(\boldsymbol{u})$, meaning the scaling of the mode shape does not affect the outcome. Consequently, $L_{\mathrm{c}}(\boldsymbol{u})$ has the $m$ zero curvature directions along the mode shapes. The pseudoinverse $[\cdot]^{\dagger}$ is applied here to bypass the singularity caused by this transformation invariance. By adopting the chain rule with matrix calculus (see Appendix A), the Hessian with and without considering the constraints is connected as

$$\nabla^2 \hat{L}_{\mathrm{c}}(\boldsymbol{u}) = \nabla^{\mathrm{T}} \hat{\boldsymbol{v}}_{\mathrm{c}}(\boldsymbol{u}) \nabla^2 \hat{L}(\boldsymbol{\theta}) \nabla \hat{\boldsymbol{v}}_{\mathrm{c}}(\boldsymbol{u}) + \left[\mathbf{I} \otimes \nabla \hat{L}(\boldsymbol{\theta})\right] \nabla^2 \hat{\boldsymbol{v}}_{\mathrm{c}}(\boldsymbol{u}) \tag{15}$$

where $\mathbf{I}$ is an identity matrix with proper size and $\otimes$ represents the Kronecker product. Typically, the second term in Eqn. (15) does not reduce to zero due to the existence of constraints. However, from the MPV computation, we know there are infinitely many parameter-sets that yield the same NLLF value, and the constraint selects only one as the MPV. Therefore, these parameter-sets are all local optimum of the NLLF, producing a zero gradient. For further clarity, a detailed proof is presented in Appendix B. This shows that the gradient of the NLLF at the renormalized local optimum (i.e., the MPV) is the product of the gradient at the local optimum before renormalization and a constant matrix. The P-EM algorithm ensures that the former is zero, so the zero-gradient of the latter can be obtained accordingly.

Now we only need to focus on the first term in Eqn. (15). The primary issue is then the selection of $\boldsymbol{u}$ and $\boldsymbol{v}_c(\boldsymbol{u})$ as they are not unique. However, the gradient of $\boldsymbol{v}_{\mathrm{c}}(\boldsymbol{u})$ is found to be the one that plays a role throughout the computation, which lies in the null space of the gradient of constraint function $\mathbf{G}(\boldsymbol{\theta})$, i.e., $\nabla \hat{\mathbf{G}}(\boldsymbol{\theta}) \nabla \hat{\boldsymbol{v}}_{\mathrm{c}}(\boldsymbol{u}) = \mathbf{0}$. Based on Appendix A, $\nabla \hat{\mathbf{G}}(\boldsymbol{\theta})$ is given by

$$\nabla \hat{\mathbf{G}}(\boldsymbol{\theta}) \in \mathbb{R}^{m \times n_\theta} = \begin{bmatrix} \nabla \hat{g}_1(\boldsymbol{\theta}) \\ \vdots \\ \nabla \hat{g}_m(\boldsymbol{\theta}) \end{bmatrix} = \begin{bmatrix} \mathbf{0} & \hat{\boldsymbol{\varphi}}_1^{\mathrm{T}} & \mathbf{0} & \mathbf{0} \\ \mathbf{0} & \mathbf{0} & \ddots & \mathbf{0} \\ \mathbf{0} & \mathbf{0} & \mathbf{0} & \hat{\boldsymbol{\varphi}}_m^{\mathrm{T}} \end{bmatrix} \tag{16}$$

By applying basic linear algebra techniques such as Gaussian-Jordan elimination [40], the null space basis $\hat{\mathbf{U}} \in \mathbb{R}^{n_\theta \times (n_\theta - m)}$ of $\nabla \hat{\mathbf{G}}(\boldsymbol{\theta})$ can be easily computed. Thus, Eqn. (14) can be updated as

$$\hat{\mathbf{C}} = \hat{\mathbf{U}} \left[\hat{\mathbf{U}}^{\mathrm{T}} \nabla^2 \hat{L}(\boldsymbol{\theta}) \hat{\mathbf{U}}\right]^{-1} \hat{\mathbf{U}}^{\mathrm{T}} \tag{17}$$

Note that the matrix within the bracket is of full rank, avoiding the potential numerical error introduced by the pseudoinverse. This form is applied in this paper to evaluate the inverse of the Hessian matrix. The remaining task is to obtain $\nabla^2 \hat{L}(\boldsymbol{\theta})$, referring to the following subsections.

## 3.2 Sparse Hessian matrix assembly

Constructing the Hessian matrix element by element during programming does not fully take advantage of the efficiency offered by matrix-based software such as MATLAB [41]. Additionally, the Hessian matrix is inherently sparse, with few non-zero entries due to the independent nature of parameters across setups. Failing to exploit this sparsity leads to significant memory overhead as the number of parameters grows, particularly when dealing with large-scale structures with tens of setups and closely-spaced modes.

To address this challenge, a sparse assembly strategy is proposed. Based on the theory in Eqn. (A.6), we first obtain the vectorized form of Eqn. (2) as

$$\mathrm{vec}(\boldsymbol{\Phi}_r) = \mathbf{K}_r \mathrm{vec}(\boldsymbol{\Phi}) \tag{18}$$

where $\mathbf{K}_r = \mathbf{I}_m \otimes \mathbf{C}_r$.The Hessian matrix at the MPV is then divided into three components for computation: the $\boldsymbol{x}^{(r)}$ part, the $\boldsymbol{\Phi}$ part, and their cross-term. Notably, since $\sum_{r=1}^{n_s} \sum_k \left[L_k^{(r)}\right]$, each of these parts can be derived from smaller, local Hessian matrices of $L_k^{(r)}$. Specifically, the first part can be directly obtained from the local Hessian without modification

$$\underbrace{\nabla^2_{\boldsymbol{x}^{(r)}} \hat{L}}_{(m+1)^2 \times (m+1)^2} = \underbrace{\nabla^2_{\boldsymbol{x}^{(r)}} \hat{L}_k^{(r)}}_{(m+1)^2 \times (m+1)^2} \tag{19}$$

For the $\boldsymbol{\Phi}$ and cross portions, the chain rule of the matrix calculus in (A.16) is applied, yielding

$$\underbrace{\nabla^2_{\boldsymbol{\Phi}\boldsymbol{\Phi}} \hat{L}}_{mn \times mn} = \sum_{r=1}^{n_s} \left[ \underbrace{\mathbf{K}_r^{\mathrm{T}}}_{mn \times mn_r} \underbrace{\nabla^2_{\boldsymbol{\Phi}_r \boldsymbol{\Phi}_r} \hat{L}_k^{(r)}}_{mn_r \times mn_r} \underbrace{\mathbf{K}_r}_{mn_r \times mn} \right] \tag{20}$$

$$\underbrace{\nabla^2_{\boldsymbol{x}^{(r)}\boldsymbol{\Phi}} \hat{L}}_{(m+1)^2 \times mn} = \sum_{r=1}^{n_s} \left[ \underbrace{\nabla^2_{\boldsymbol{x}^{(r)}\boldsymbol{\Phi}_r} \hat{L}_k^{(r)}}_{(m+1)^2 \times mn_r} \underbrace{\mathbf{K}_r}_{mn_r \times mn} \right] = \underbrace{\nabla^2_{\boldsymbol{\Phi}\boldsymbol{x}^{(r)}} \hat{L}}_{mn \times (m+1)^2} \tag{21}$$

For these parts, we can bypass the direct multiplication with the Kronecker product term $\mathbf{K}_r$ by leveraging the sparsity and known structure of $\mathbf{C}_r$. The term $\nabla^2_{\boldsymbol{\Phi}_r \boldsymbol{\Phi}_r} \hat{L}_k^{(r)}$ is separated into $m^2$ block matrices in $\mathbb{R}^{n_r \times n_r}$ and recorded as $\widehat{\mathbf{H}}_{\boldsymbol{\Phi}_r}^{(ij)}$, where $i$ and $j$ range from 1 to $m$. Likewise, $\nabla^2_{\boldsymbol{\Phi}\boldsymbol{\Phi}} \hat{L}$ is decomposed into $m^2$ smaller matrices of size $n \times n$ using the same division, recorded as $\widehat{\mathbf{H}}_{\boldsymbol{\Phi}}^{(ij)}$. In this way, Eqn. (20) is rewritten as

$$\begin{bmatrix} \widehat{\mathbf{H}}_{\boldsymbol{\Phi}}^{(11)} & \cdots & \widehat{\mathbf{H}}_{\boldsymbol{\Phi}}^{(1m)} \\ \vdots & \ddots & \vdots \\ \widehat{\mathbf{H}}_{\boldsymbol{\Phi}}^{(m1)} & \cdots & \widehat{\mathbf{H}}_{\boldsymbol{\Phi}}^{(mm)} \end{bmatrix} = \sum_{r=1}^{n_s} \left( \begin{bmatrix} \mathbf{C}_r^{\mathrm{T}} & \cdots & \mathbf{C}_r^{\mathrm{T}} \\ \vdots & \ddots & \vdots \\ \mathbf{C}_r^{\mathrm{T}} & \cdots & \mathbf{C}_r^{\mathrm{T}} \end{bmatrix} \begin{bmatrix} \widehat{\mathbf{H}}_{\boldsymbol{\Phi}_r}^{(11)} & \cdots & \widehat{\mathbf{H}}_{\boldsymbol{\Phi}_r}^{(1m)} \\ \vdots & \ddots & \vdots \\ \widehat{\mathbf{H}}_{\boldsymbol{\Phi}_r}^{(m1)} & \cdots & \widehat{\mathbf{H}}_{\boldsymbol{\Phi}_r}^{(mm)} \end{bmatrix} \begin{bmatrix} \mathbf{C}_r & \cdots & \mathbf{C}_r \\ \vdots & \ddots & \vdots \\ \mathbf{C}_r & \cdots & \mathbf{C}_r \end{bmatrix} \right) \tag{22}$$

Note that there is only one entry of 1 in each row, while the columns may contain more than one 1 in the selection matrix $\mathbf{C}_r$. Define the position index of the unique 1 in each row as $\boldsymbol{\tau}^{(r)} = \left[\tau_1^{(r)}, \tau_2^{(r)}, \cdots, \tau_{n_r}^{(r)}\right]^{\mathrm{T}} \in \mathbb{R}^{n_r \times 1}$. Let $\boldsymbol{e}_{\tau_u^{(r)}} \in \mathbb{R}^{n \times 1}$ $(u = 1, 2, \cdots, n_r)$ be a binary vector with only one 1 element corresponding to the subscript $\tau_u^{(r)}$. As a result, $\mathbf{C}_r = \left[\boldsymbol{e}_{\tau_1^{(r)}}, \boldsymbol{e}_{\tau_2^{(r)}}, \cdots, \boldsymbol{e}_{\tau_{n_r}^{(r)}}\right]^{\mathrm{T}}$ and we need only to compute each small part

$$\widehat{\mathbf{H}}_{\boldsymbol{\Phi}}^{(ij)} = \sum_{r=1}^{n_s} \mathbf{C}_r^{\mathrm{T}} \widehat{\mathbf{H}}_{\boldsymbol{\Phi}_r}^{(ij)} \mathbf{C}_r = \sum_{r=1}^{n_s} \widehat{\mathbf{H}}_{\boldsymbol{\Phi}_r}^{(ij)}(u,v) \boldsymbol{e}_{\tau_u^{(r)}} \boldsymbol{e}_{\tau_v^{(r)}}^{\mathrm{T}} \tag{23}$$

where $u$ and $v$ range from 1 to $n_r$. The $(u,v)$-th element of $\widehat{\mathbf{H}}_{\boldsymbol{\Phi}_r}^{(ij)}$ is mapped into the $\left(\tau_u^{(r)}, \tau_v^{(r)}\right)$ position of $\widehat{\mathbf{H}}_{\boldsymbol{\Phi}}^{(ij)}$, making it possible to compute the above equation by simple assignment operation.

Similarly, divide $\nabla^2_{\boldsymbol{x}^{(r)}\boldsymbol{\Phi}} \hat{L}$ into $m$ blocks $\widehat{\mathbf{H}}_{\boldsymbol{x}^{(r)}\boldsymbol{\Phi}}^{(i)} \in \mathbb{R}^{(m+1)^2 \times n}$ and decompose $\widehat{\mathbf{H}}_{\boldsymbol{x}^{(r)}\boldsymbol{\Phi}}^{(i)}$ as $\widehat{\mathbf{H}}_{\boldsymbol{x}^{(r)}\boldsymbol{\Phi}}^{(j)} \in \mathbb{R}^{(m+1)^2 \times n_r}$ accordingly. A memory-saving implementation of Eqn. (21) is given by

$$\widehat{\mathbf{H}}_{\boldsymbol{x}^{(r)}\boldsymbol{\Phi}}^{(i)} = \sum_{r=1}^{n_s} \widehat{\mathbf{H}}_{\boldsymbol{x}^{(r)}\boldsymbol{\Phi}_r}^{(i)} \mathbf{C}_r = \sum_{r=1}^{n_s} \widehat{\mathbf{H}}_{\boldsymbol{x}^{(r)}\boldsymbol{\Phi}_r}^{(i)}(:,v) \boldsymbol{e}_{\tau_v^{(r)}}^{\mathrm{T}} \tag{24}$$

where $\widehat{\mathbf{H}}_{\boldsymbol{x}^{(r)}\boldsymbol{\Phi}_r}^{(i)}(:,v)$ denotes the $v$-th column of $\widehat{\mathbf{H}}_{\boldsymbol{x}^{(r)}\boldsymbol{\Phi}_r}^{(i)}$. The term $\widehat{\mathbf{H}}_{\boldsymbol{x}^{(r)}\boldsymbol{\Phi}}^{(i)}$ is seen to be obtained by rearranging the columns of $\widehat{\mathbf{H}}_{\boldsymbol{x}^{(r)}\boldsymbol{\Phi}_r}^{(i)}$, and summing them up. The final step of the proposed method involves obtaining the small matrix components $\nabla^2_{\boldsymbol{x}^{(r)}} \hat{L}_k^{(r)}$, $\nabla^2_{\boldsymbol{\Phi}_r\boldsymbol{\Phi}_r} \hat{L}_k^{(r)}$, and $\nabla^2_{\boldsymbol{x}^{(r)}\boldsymbol{\Phi}_r} \hat{L}_k^{(r)}$, as illustrated in the next subsection.

## 3.3 Indirect second-order derivative computation

For getting the essential components of the Hessian matrix, Zhu and Au [34] proposed a method for directly manipulating the NLLF. The derivatives w. r. t. parameters other than the mode shape are computed in an element-wise manner, posing challenges for both derivation and implementation. A key difficulty arises from the nonlinear coupling of parameters in the NLLF. When computing the second-order derivatives, the chain rule must be applied multiple times, leading to a large number of intermediate quantities and complicating the process further.

An alternative approach leverages the decoupled structure in the complete-data NLLF, as applied in the MPV computation. In this context, Louis' identity [42] provides a link between the Hessian of the NLLF before and after introducing the latent variable. This technique has been successfully implemented in the single-setup Bayesian FFT method and is at least an order of magnitude faster than directly working with NLLF [43]. To extend this approach to the multi-setup case, we first construct a 'local' MPV set, $\widehat{\boldsymbol{\theta}}^{(r)} = \left[\widehat{\boldsymbol{x}}^{(r)}; \mathrm{vec}\left[\widehat{\boldsymbol{\Phi}}_r\right]\right]$, where $\widehat{\boldsymbol{\Phi}}_r$ is obtained by $\mathbf{C}_r \widehat{\boldsymbol{\Phi}}$. Using this, the local Hessian is computed based on the Fisher's identity and Louis' identity [43] as

$$\nabla \hat{L}_k^{(r)} = \nabla \hat{Q}_k^{(r)} \tag{25}$$

$$\nabla^2 \hat{L}_r = \sum\nolimits_k \nabla^{\mathrm{T}} \hat{L}_k^{(r)} \nabla \hat{L}_k^{(r)} + \sum\nolimits_k \nabla^2 \hat{Q}_k^{(r)} - \sum\nolimits_k \mathrm{E}_{\hat{\boldsymbol{\theta}}^{(r)}} \left[ \nabla^{\mathrm{T}} \hat{\mathcal{L}}_k^{(r)} \nabla \hat{\mathcal{L}}_k^{(r)} \right] \tag{26}$$

where the dependence of $\hat{Q}_k^{(r)}$ and $\hat{\mathcal{L}}_k^{(r)}$ on $\hat{\boldsymbol{\theta}}^{(r)}$ is omitted for simplicity. The Hessian is seen to be separated into three parts, which are related to the gradient, Hessian, and expectation of the complete-data NLLF. Because the parameters are partially decoupled in this function, the process of taking derivatives is simplified. For a more detailed step-by-step derivation, refer to Zhu and Li [43], while the main results are presented in **Table 1** and **Table 2**.

**Table 1.** First-order derivatives of $\hat{Q}_k^{(r)}$. (adapted from Zhu and Li [43])

| Notation | Result |
|---|---|
| $\nabla_{\boldsymbol{f}^{(r)}} \hat{Q}_k^{(r)}$ | $2\mathrm{Re}\left[-\left(\frac{2\hat{\boldsymbol{f}}^{(r)\mathrm{T}}}{\mathrm{f}_k^{(r)2}} + \frac{2\mathbf{i}\hat{\boldsymbol{\zeta}}^{(r)\mathrm{T}}}{\mathrm{f}_k^{(r)}}\right)\left(\widehat{\mathbf{W}}_k^{(r)}\left[\hat{\mathbf{h}}_k^{(r)*}\right]^{-1}\left[\hat{\mathbf{S}}^{(r)}\right]^{-1} \odot \mathbf{I}_m\right)\right] + \left(\frac{4\hat{\boldsymbol{f}}^{(r)\mathrm{T}}}{\mathrm{f}_k^{(r)2}} - \frac{4\hat{\boldsymbol{f}}^{(r)3\mathrm{T}}}{\mathrm{f}_k^{(r)4}} - \frac{8\hat{\boldsymbol{f}}^{(r)\mathrm{T}} \odot \hat{\boldsymbol{\zeta}}^{(r)2\mathrm{T}}}{\mathrm{f}_k^{(r)2}}\right)\widehat{\mathbf{D}}_k^{(r)}$ |
| $\nabla_{\boldsymbol{\zeta}^{(r)}} \hat{Q}_k^{(r)}$ | $2\mathrm{Re}\left[-\frac{2\mathbf{i}}{\mathrm{f}_k^{(r)}}\hat{\boldsymbol{f}}^{(r)\mathrm{T}}\left(\widehat{\mathbf{W}}_k^{(r)}\left[\hat{\mathbf{h}}_k^{(r)*}\right]^{-1}\left[\hat{\mathbf{S}}^{(r)}\right]^{-1} \odot \mathbf{I}_m\right)\right] - \frac{8\hat{\boldsymbol{f}}^{(r)2\mathrm{T}} \odot \hat{\boldsymbol{\zeta}}^{\mathrm{T}}\widehat{\mathbf{D}}_k^{(r)}}{\mathrm{f}_k^{(r)2}}$ |
| $\nabla_{\mathbf{S}^{(r)}} \hat{Q}_k^{(r)}$ | $\mathrm{vec}^{\mathrm{T}}\left(-\left[\hat{\mathbf{S}}^{(r)}\right]^{-\mathrm{T}}\left[\hat{\mathbf{h}}_k^{(r)*}\right]^{-1}\widehat{\mathbf{W}}_k^{(r)\mathrm{T}}\left[\hat{\mathbf{h}}_k^{(r)}\right]^{-1}\left[\hat{\mathbf{S}}^{(r)}\right]^{-\mathrm{T}} + \left[\hat{\mathbf{S}}^{(r)}\right]^{-\mathrm{T}}\right)$ |
| $\nabla_{S_e^{(r)}} \hat{Q}_k^{(r)}$ | $1/\left(n_r S_{\mathrm{e}}^{(r)}\right) - \hat{\boldsymbol{\mathcal{F}}}_k^{(r)\mathrm{H}}\hat{\boldsymbol{\mathcal{F}}}_k^{(r)}/S_{\mathrm{e}}^{(r)2} + \left(\hat{\boldsymbol{\mathcal{F}}}_k^{(r)\mathrm{H}}\hat{\boldsymbol{\Phi}}_r\hat{\boldsymbol{w}}_k^{(r)} + \hat{\boldsymbol{w}}_k^{(r)\mathrm{H}}\hat{\boldsymbol{\Phi}}_r^{\mathrm{T}}\hat{\boldsymbol{\mathcal{F}}}_k^{(r)}\right)/S_{\mathrm{e}}^{(r)2} - \mathrm{tr}\left(\widehat{\mathbf{W}}_k^{(r)}\hat{\boldsymbol{\Phi}}_r^{\mathrm{T}}\hat{\boldsymbol{\Phi}}_r\right)/S_{\mathrm{e}}^{(r)2}$ |
| $\nabla_{\boldsymbol{\Phi}_r} \hat{Q}_k^{(r)}$ | $1/S_{\mathrm{e}}^{(r)}\,\mathrm{vec}^{\mathrm{T}}\left[-\hat{\boldsymbol{\mathcal{F}}}_k^{(r)*}\hat{\boldsymbol{w}}_k^{(r)\mathrm{T}} - \hat{\boldsymbol{\mathcal{F}}}_k^{(r)}\hat{\boldsymbol{w}}_k^{(r)\mathrm{H}} + \hat{\boldsymbol{\Phi}}_r\left(\widehat{\mathbf{W}}_k^{(r)} + \widehat{\mathbf{W}}_k^{(r)\mathrm{T}}\right)\right]$ |

Note: $\mathbf{W}_k^{(r)} = \hat{\boldsymbol{w}}_k^{(r)}\hat{\boldsymbol{w}}_k^{(r)\mathrm{H}} + \hat{\boldsymbol{\Sigma}}_k^{(r)}$, $\widehat{\mathbf{D}}_k^{(r)} = \hat{\mathbf{h}}_k^{(r)}\hat{\mathbf{h}}_k^{(r)*}$, $\boldsymbol{f}^{(r)p} = \underbrace{\boldsymbol{f}^{(r)} \odot \cdots \odot \boldsymbol{f}^{(r)}}_{p}$, and $\boldsymbol{\zeta}^{(r)p} = \underbrace{\boldsymbol{\zeta}^{(r)} \odot \cdots \odot \boldsymbol{\zeta}^{(r)}}_{p}$.

**Table 2.** Second-order derivatives of the $\hat{Q}_k^{(r)}$. (adapted from Zhu and Li [43]).

| Notation | Result |
|---|---|
| $\nabla^2_{\boldsymbol{f}^{(r)}\boldsymbol{f}^{(r)}} \hat{Q}_k$ | $2\mathrm{Re}\left[\mathrm{diag}\left(\frac{2}{\mathrm{f}_k^{(r)2}}\hat{\boldsymbol{f}}^{(r)} + \frac{2\mathbf{i}}{\mathrm{f}_k^{(r)}}\hat{\boldsymbol{\zeta}}^{(r)}\right)\left[\hat{\mathbf{S}}^{(r)}\right]^{-\mathrm{T}} \odot \widehat{\mathbf{W}}_k^{(r)}\mathrm{diag}\left(\frac{2}{\mathrm{f}_k^{(r)2}}\hat{\boldsymbol{f}}^{(r)} - \frac{2\mathbf{i}}{\mathrm{f}_k^{(r)}}\hat{\boldsymbol{\zeta}}^{(r)}\right)\right]$ <br> $-2\mathrm{Re}\left[\frac{2}{\mathrm{f}_k^{(r)2}}\left(\widehat{\mathbf{W}}_k^{(r)}\left[\hat{\mathbf{h}}_k^{(r)*}\right]^{-1}\left[\hat{\mathbf{S}}^{(r)}\right]^{-1} \odot \mathbf{I}_{\mathrm{m}}\right)\right] + \widehat{\mathbf{D}}_k^{(r)}\mathrm{diag}\left(\frac{4}{\mathrm{f}_k^{(r)2}} - \frac{12\hat{\boldsymbol{f}}^{(r)2}}{\mathrm{f}_k^{(r)4}} - \frac{8\hat{\boldsymbol{\zeta}}^{(r)2}}{\mathrm{f}_k^{(r)2}}\right) +$ <br> $\widehat{\mathbf{D}}_k^{(r)2}\mathrm{diag}\left(\frac{16\hat{\boldsymbol{f}}^{(r)2}}{\mathrm{f}_k^{(r)4}} - \frac{32\hat{\boldsymbol{f}}^{(r)4}}{\mathrm{f}_k^{(r)6}} - \frac{64\hat{\boldsymbol{\zeta}}^{(r)2} \odot \hat{\boldsymbol{f}}^{(r)2}}{\mathrm{f}_k^{(r)4}} + \frac{16\hat{\boldsymbol{f}}^{(r)6}}{\mathrm{f}_k^{(r)8}} + \frac{64\hat{\boldsymbol{\zeta}}^{(r)2} \odot \hat{\boldsymbol{f}}^{(r)4}}{\mathrm{f}_k^{(r)6}} + \frac{64\hat{\boldsymbol{\zeta}}^{(r)4} \odot \hat{\boldsymbol{f}}^{(r)4}}{\mathrm{f}_k^{(r)4}}\right)$ |
| $\nabla^2_{\boldsymbol{\zeta}^{(r)}\boldsymbol{\zeta}^{(r)}}\hat{Q}_k$ | $\frac{64\hat{\mathbf{D}}_k^{(r)2}\mathrm{diag}(\hat{\boldsymbol{f}}^{(r)4}\odot\hat{\boldsymbol{\zeta}}^{(r)2})}{\mathrm{f}_k^{(r)4}}-\frac{8\hat{\mathbf{D}}_k^{(r)}\mathrm{diag}(\hat{\boldsymbol{f}}^{(r)2})}{\mathrm{f}_k^{(r)2}}-2\mathrm{Re}\left[\mathrm{diag}\left(\frac{2\mathbf{i}\hat{\boldsymbol{f}}^{(r)}}{\mathrm{f}_k^{(r)}}\right)\left[\hat{\mathbf{S}}^{(r)}\right]^{-\mathrm{T}}\odot\widehat{\mathbf{W}}_k^{(r)}\mathrm{diag}\left(\frac{2\mathbf{i}\hat{\boldsymbol{f}}^{(r)}}{\mathrm{f}_k^{(r)}}\right)\right]$ |
| $\nabla^2_{\mathbf{S}^{(r)}\mathbf{S}^{(r)}}\hat{Q}_k$ | $\mathbf{K}_{\mathrm{mm}}\left[\hat{\mathbf{S}}^{(r)}\right]^{-\mathrm{T}}\left[\hat{\mathbf{h}}_k^{(r)*}\right]^{-1}\widehat{\mathbf{W}}_k^{(r)\mathrm{T}}\left[\hat{\mathbf{h}}_k^{(r)}\right]^{-1}\left[\hat{\mathbf{S}}^{(r)}\right]^{-\mathrm{T}}\otimes\left[\hat{\mathbf{S}}^{(r)}\right]^{-\mathrm{T}}$ <br> $+\mathbf{K}_{\mathrm{mm}}\left[\hat{\mathbf{S}}^{(r)}\right]^{-\mathrm{T}}\otimes\left(\left[\hat{\mathbf{S}}^{(r)}\right]^{-1}\left[\hat{\mathbf{h}}_k^{(r)}\right]^{-1}\widehat{\mathbf{W}}_k^{(r)}\left[\hat{\mathbf{h}}_k^{(r)*}\right]^{-1}\left[\hat{\mathbf{S}}^{(r)}\right]^{-1}\right)-\mathbf{K}_{\mathrm{mm}}\left[\hat{\mathbf{S}}^{(r)}\right]^{-\mathrm{T}}\otimes\left[\hat{\mathbf{S}}^{(r)}\right]^{-1}$ |
| $\nabla^2_{S_e^{(r)}S_e^{(r)}}\hat{Q}_k$ | $-1/\left(n_r S_{\mathrm{e}}^{(r)2}\right)+2\left[\hat{\boldsymbol{\mathcal{F}}}_k^{(r)\mathrm{H}}\hat{\boldsymbol{\mathcal{F}}}_k^{(r)}-\mathbf{2}\mathrm{tr}\left[\hat{\boldsymbol{\Phi}}_r\mathrm{Re}\left(\hat{\boldsymbol{w}}_k^{(r)}\hat{\boldsymbol{\mathcal{F}}}_k^{(r)\mathrm{H}}\right)\right]+\mathrm{tr}\left[\hat{\boldsymbol{\Phi}}_r\mathrm{Re}\left(\widehat{\mathbf{W}}_k^{(r)}\right)\hat{\boldsymbol{\Phi}}_r^{\mathrm{T}}\right]\right]/S_{\mathrm{e}}^{(r)3}$ |
| $\nabla^2_{\boldsymbol{\Phi}_r\boldsymbol{\Phi}_r}\hat{Q}_k$ | $\left(\widehat{\mathbf{W}}_k^{(r)}+\widehat{\mathbf{W}}_k^{(r)\mathrm{T}}\right)\otimes\mathbf{I}_{\mathrm{m}}/S_{\mathrm{e}}^{(r)}$ |
| $\nabla^2_{\boldsymbol{f}^{(r)}\boldsymbol{\zeta}^{(r)}}\hat{Q}_k$ | $-2\mathrm{Re}\left[\frac{2\mathbf{i}\widehat{\mathbf{W}}_k^{(r)}\left[\hat{\mathbf{h}}_k^{(r)*}\right]^{-1}\left[\hat{\mathbf{S}}^{(r)}\right]^{-1}\odot\mathbf{I}_{\mathrm{m}}}{\mathrm{f}_k^{(r)}}-\mathrm{diag}\left(\frac{2\hat{\boldsymbol{f}}^{(r)}}{\mathrm{f}_k^{(r)2}}+\frac{2\mathbf{i}\hat{\boldsymbol{\zeta}}^{(r)}}{\mathrm{f}_k^{(r)}}\right)\left[\hat{\mathbf{S}}^{(r)}\right]^{-\mathrm{T}}\odot\widehat{\mathbf{W}}_k^{(r)}\mathrm{diag}\left(\frac{2\mathbf{i}\hat{\boldsymbol{f}}^{(r)}}{\mathrm{f}_k^{(r)}}\right)\right]$ <br> $+\hat{\mathbf{D}}_k^{(r)2}\mathrm{diag}\left(-\frac{32\hat{\boldsymbol{f}}^{(r)3}\odot\hat{\boldsymbol{\zeta}}^{(r)}}{\mathrm{f}_k^{(r)4}}+\frac{32\hat{\boldsymbol{f}}^{(r)5}\odot\hat{\boldsymbol{\zeta}}^{(r)}}{\mathrm{f}_k^{(r)6}}+\frac{64\hat{\boldsymbol{f}}^{(r)3}\odot\hat{\boldsymbol{\zeta}}^{(r)3}}{\mathrm{f}_k^{(r)4}}\right)-\hat{\mathbf{D}}_k^{(r)}\mathrm{diag}\left(\frac{16\hat{\boldsymbol{f}}^{(r)}\odot\hat{\boldsymbol{\zeta}}^{(r)}}{\mathrm{f}_k^{(r)2}}\right)$ |
| $\nabla^2_{\boldsymbol{f}^{(r)}\mathbf{S}^{(r)}}\hat{Q}_k$ | $\mathrm{diag}\left(\frac{2\hat{\boldsymbol{f}}^{(r)}}{\mathrm{f}_k^{(r)2}}+\frac{2\mathbf{i}\hat{\boldsymbol{\zeta}}^{(r)}}{\mathrm{f}_k^{(r)}}\right)\mathbf{R}_{\mathrm{m}}\left[\mathbf{I}_{\mathrm{m}}\otimes\left(\widehat{\mathbf{W}}_k^{(r)}\left[\hat{\mathbf{h}}_k^{(r)*}\right]^{-1}\right)\right]\left(\left[\hat{\mathbf{S}}^{(r)}\right]^{-\mathrm{T}}\otimes\left[\hat{\mathbf{S}}^{(r)}\right]^{-1}\right)$ <br> $+\mathrm{diag}\left(\frac{2\hat{\boldsymbol{f}}^{(r)}}{\mathrm{f}_k^{(r)2}}-\frac{2\mathbf{i}\hat{\boldsymbol{\zeta}}^{(r)}}{\mathrm{f}_k^{(r)}}\right)\mathbf{R}_{\mathrm{m}}\left[\mathbf{I}_{\mathrm{m}}\otimes\left(\widehat{\mathbf{W}}_k^{(r)}\left[\hat{\mathbf{h}}_k^{(r)}\right]^{-1}\right)\right]\left(\left[\hat{\mathbf{S}}^{(r)}\right]^{-\mathrm{T}}\otimes\left[\hat{\mathbf{S}}^{(r)}\right]^{-1}\right)\mathbf{K}_{\mathrm{mm}}$ |
| $\nabla^2_{\boldsymbol{\zeta}^{(r)}\mathbf{S}^{(r)}}\hat{Q}_k$ | $\mathrm{diag}\left(\frac{2\mathbf{i}\hat{\boldsymbol{f}}^{(r)}}{\mathrm{f}_k^{(r)}}\right)\mathbf{R}_{\mathrm{m}}\left(\mathbf{I}_{\mathrm{m}}\otimes\left(\widehat{\mathbf{W}}_k^{(r)}\left[\hat{\mathbf{h}}_k^{(r)*}\right]^{-1}\right)\right)\left(\left[\hat{\mathbf{S}}^{(r)}\right]^{-\mathrm{T}}\otimes\left[\hat{\mathbf{S}}^{(r)}\right]^{-1}\right)$ <br> $-\mathrm{diag}\left(\frac{2\mathbf{i}\hat{\boldsymbol{f}}^{(r)}}{\mathrm{f}_k^{(r)}}\right)\mathbf{R}_{\mathrm{m}}\left(\mathbf{I}_{\mathrm{m}}\otimes\left(\widehat{\mathbf{W}}_k^{(r)}\left[\hat{\mathbf{h}}_k^{(r)}\right]^{-1}\right)\right)\left(\left[\hat{\mathbf{S}}^{(r)}\right]^{-\mathrm{T}}\otimes\left[\hat{\mathbf{S}}^{(r)}\right]^{-1}\right)\mathbf{K}_{\mathrm{mm}}$ |
| $\nabla^2_{\boldsymbol{\Phi}_r S_e^{(r)}}\hat{Q}_k$ | $-\mathrm{vec}\left[-\hat{\boldsymbol{\mathcal{F}}}_k^{(r)*}\hat{\boldsymbol{w}}_k^{(r)\mathrm{T}}-\hat{\boldsymbol{\mathcal{F}}}_k^{(r)}\hat{\boldsymbol{w}}_k^{(r)\mathrm{H}}+\hat{\boldsymbol{\Phi}}_r\left(\widehat{\mathbf{W}}_k^{(r)}+\widehat{\mathbf{W}}_k^{(r)\mathrm{T}}\right)\right]/S_{\mathrm{e}}^{(r)2}$ |

Note: $\nabla^2_{\mathbf{xy}}\hat{Q}_k \triangleq \frac{\partial}{\partial \boldsymbol{y}}\frac{\partial Q}{\partial \boldsymbol{x}^{\mathrm{T}}}\|_{\boldsymbol{\theta}=\hat{\boldsymbol{\theta}}}$, where $\boldsymbol{x}$ and $\boldsymbol{y}$ are a parameter in $\boldsymbol{\theta}^{(r)}$.

Here, the commutation matrix $\mathbf{K}_{\mathrm{mm}} \in \mathbb{R}^{m\times m^2}$ is defined in Appendix A; Another permutation matrix, $\mathbf{R}_{\mathrm{m}} \in \mathbb{R}^{m\times m^2}$, connects the diagonal entries of an arbitrary diagonal matrix $\mathbf{A} \in \mathbb{R}^{m\times m}$ to its vectorized form, such that $\mathbf{A}_{ii} = \mathbf{R}_{\mathrm{m}}\mathrm{vec}(\mathbf{A})$, where $\mathbf{A}_{ii} \in \mathbb{R}^{m\times 1}$ represents the vector of diagonal entries of $\mathbf{A}$.

Based on these results, the first and second terms in Eqn. (26) can be obtained. To calculate the final term, i.e., the expected product of the gradient of the complete-data NLLF, the first-order derivatives should be obtained first. This is the byproduct of the results in **Table 1**, as expectation and differentiation are both linear operators and thus commutative. By replacing the moments $\hat{\boldsymbol{w}}_k^{(r)}$ and $\widehat{\mathbf{W}}_k^{(r)}$ with $\boldsymbol{\eta}_k^{(r)}$ and $\boldsymbol{\eta}_k^{(r)}\boldsymbol{\eta}_k^{(r)\mathrm{H}}$, the expression for $\nabla\hat{\mathcal{L}}_k^{(r)}$ can be derived. Since the expectation is taken w. r. t. the Gaussian distribution $p\left(\boldsymbol{\eta}_k^{(r)}\middle|\hat{\boldsymbol{\mathcal{F}}}_k^{(r)},\boldsymbol{\theta}\right)$, the moments are available in analytical form, as shown in Eqns. (10) and (11). With the expectation terms resolved, the local Hessian parts required for Eqns. (19), (22), and (23) are now fully computed.

In conclusion, the pseudocode of the fast PCM computation method is outlined below.

**Algorithm 1.** EM-based fast PCM computation in multi-setup Bayesian FFT method.

---

**Input:** FFT data within the selected band for each setup $\{\hat{\mathcal{F}}_k^{(r)}\}$;

Selection matrix for each setup $\mathbf{C}_r$;

MPV of modal parameters $\hat{\boldsymbol{\theta}}$.

**Output:** Posterior covariance matrix $\hat{\mathbf{C}}$.

% Initialization.

Calculate the MPV of the local mode shape $\hat{\boldsymbol{\Phi}}_r = \mathbf{C}_r \hat{\boldsymbol{\Phi}}$;

Generates all zero sparse matrices $\hat{\mathbf{H}}_{\boldsymbol{\Phi}}$ and $\hat{\mathbf{H}}_{\boldsymbol{x}^{(r)}\boldsymbol{\Phi}}$;

% Assemble Hessian matrix. See Sections 3.3 and 3.2.

**for** $r = 1{:}n_s$, **do**

  Compute $\hat{\mathrm{E}}_1^{(r)} = \sum_k \nabla^{\mathrm{T}} \hat{Q}_k^{(r)} \nabla \hat{Q}_k^{(r)}$;  % See **Table 1**.

  Compute $\hat{\mathrm{E}}_2^{(r)} = \sum_k \nabla^2 \hat{Q}_k^{(r)}$;  % See **Table 2**.

  Compute $\hat{\mathrm{E}}_3^{(r)} = -\sum_k \mathrm{E}_{\hat{\boldsymbol{\theta}}^{(r)}} \left[\nabla^{\mathrm{T}} \hat{\mathcal{L}}_k^{(r)} \nabla \hat{\mathcal{L}}_k^{(r)}\right]$;  % See **Table 1** and Eqns. (19), (22), and (23).

  $\hat{\mathbf{H}}_{\boldsymbol{\theta}^{(r)}} = \hat{\mathrm{E}}_1^{(r)} + \hat{\mathrm{E}}_2^{(r)} + \hat{\mathrm{E}}_3^{(r)}$;  % See Eqn. (26).

  Decompose $\hat{\mathbf{H}}_{\boldsymbol{\theta}^{(r)}}$ as $\begin{bmatrix} \hat{\mathbf{H}}_{\boldsymbol{x}^{(r)}} & \hat{\mathbf{H}}_{\boldsymbol{x}^{(r)}\boldsymbol{\Phi}_r} \\ \hat{\mathbf{H}}_{\boldsymbol{x}^{(r)}\boldsymbol{\Phi}_r}^{\mathrm{T}} & \hat{\mathbf{H}}_{\boldsymbol{\Phi}_r} \end{bmatrix}$.

  $\boldsymbol{\tau}^{(r)} = [\mathrm{find}(\mathbf{C}_r^{\mathrm{T}})]^{\mathrm{T}} \in \mathbb{R}^{n_r \times 1}$;  % Find the nonzero element in each row of $\mathbf{C}_r$

  $\hat{\mathbf{H}}_{\boldsymbol{\Phi}}(\boldsymbol{\tau}^{(r)}, \boldsymbol{\tau}^{(r)}) = \hat{\mathbf{H}}_{\boldsymbol{\Phi}}(\boldsymbol{\tau}^{(r)}, \boldsymbol{\tau}^{(r)}) + \hat{\mathbf{H}}_{\boldsymbol{\Phi}_r}$;  % See Eqn. (23)

  $\hat{\mathbf{H}}_{\boldsymbol{x}^{(r)}\boldsymbol{\Phi}}(:, \boldsymbol{\tau}^{(r)}) = \hat{\mathbf{H}}_{\boldsymbol{x}^{(r)}\boldsymbol{\Phi}}(:, \boldsymbol{\tau}^{(r)}) + \hat{\mathbf{H}}_{\boldsymbol{x}^{(r)}\boldsymbol{\Phi}_r}$;  % See Eqn. (24)

**end for**

Assemble $\hat{\mathbf{H}}_{\boldsymbol{\theta}} = \begin{bmatrix} \mathrm{blkdiag}\left(\{\hat{\mathbf{H}}_{\boldsymbol{x}^{(r)}}\}_{r=1}^{n_s}\right) & \hat{\mathbf{H}}_{\boldsymbol{x}^{(r)}\boldsymbol{\Phi}} \\ \hat{\mathbf{H}}_{\boldsymbol{x}^{(r)}\boldsymbol{\Phi}}^{\mathrm{T}} & \hat{\mathbf{H}}_{\boldsymbol{\Phi}} \end{bmatrix}$;  % blkdiag(·) denotes creating a block diagonal matrix

% Solve mode shape norm constraints singularity. See Section. 3.1.

Solve $\nabla \hat{\mathbf{G}}(\boldsymbol{\theta}) \hat{\mathbf{U}} = \mathbf{0}$ to get the nullspace basis $\hat{\mathbf{U}}$.  % See Eqn. (16)

Return PCM $\hat{\mathbf{C}} = \hat{\mathbf{U}}\left[\hat{\mathbf{U}}^{\mathrm{T}} \hat{\mathbf{H}}_{\boldsymbol{\theta}} \hat{\mathbf{U}}\right]^{-1} \hat{\mathbf{U}}^{\mathrm{T}}$.

---

# 4 Synthetic example verification

In this section, the proposed method is applied in a synthetic example to both validate the correctness and assess the key factors influencing computational efficiency.

## 4.1 Problem setting

The target structure is an eight-story synthetic shear-frame model, with a total of 34 points actively measured, as illustrated in **Fig. 1** a). Notably, Points (2) and (4), located at the center of the top two sides, are intentionally additional to facilitate the configuration of reference and rover DoFs. This design ensures that all available sensors are fully utilized, avoiding the inefficiency of underused resources during each test. The right subfigure in **Fig. 1** a) highlights the corresponding channel directions and orientations on a typical floor, which is essential for implementing various multi-setup vibration tests. In **Fig. 1** b), the 'true' modal parameters of the structure used for generating the data are presented, including the natural frequencies, damping ratios, and mode shapes for three modes: translational in the X direction (TX), translational in the Y direction (TY), and rotational (R). The modal frequencies and corresponding damping ratios are set as $f_1 = 4.20$ Hz, $\zeta_1 = 1.0\%$, $f_2 = 4.25$ Hz, $\zeta_2 = 1.5\%$, and $f_3 = 4.40$ Hz, $\zeta_3 = 2.0\%$, respectively. For this example, the modal excitation is assumed to be stationary Gaussian white noise with an auto-PSD of $1\ (\mu g)^2/\text{Hz}$. The cross-PSD between the first- and second-order modal excitation is modeled as $e^{i\pi/4}$, where $e$ is the base of the natural logarithm, and the expression represents a phase shift of $\pi/4$ radians in the complex plane. The other cross-PSD terms are set to zero, simplifying the spectral interaction among the modes. Moreover, the prediction error is also modeled as stationary Gaussian white noise with a PSD of $10\ (\mu g)^2/\text{Hz}$, to account for inherent uncertainties in the response data.

Using these modal parameters, synthetic response data is generated for all 68 DoFs over 4 hours, sampled at 100 Hz. To simulate the multi-setup test commonly performed in OMA, subsets of response data are systematically selected over different periods. This allows a comprehensive analysis of the vibrational response of the structure with a limited number of biaxial sensors. By repositioning sensors across multiple setups, it ensures all modes are adequately captured and recorded for analysis.

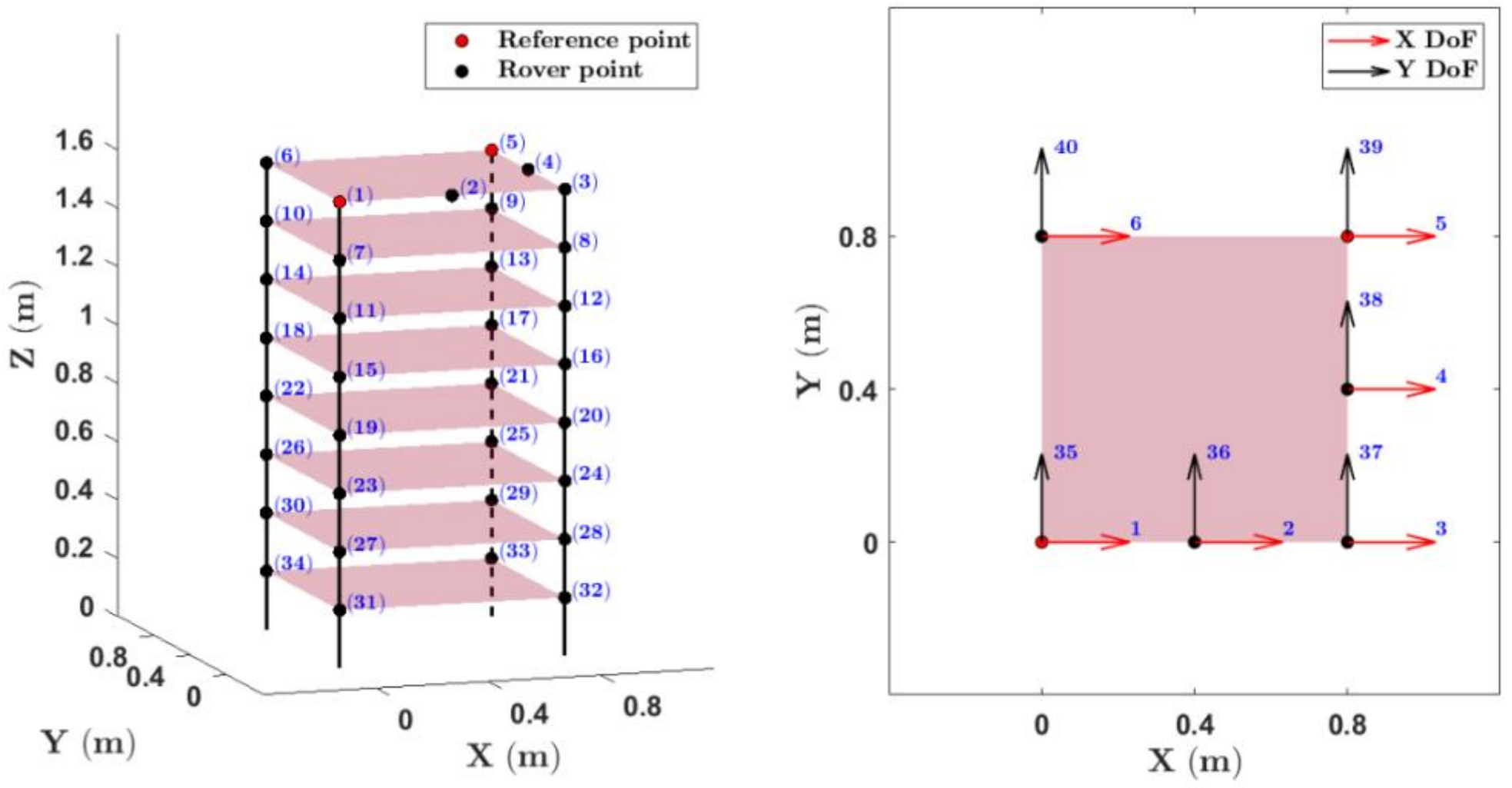

a) Measured point and channel numbering. Numbers inside and outside the parentheses denote DoF and channel numbering, respectively.

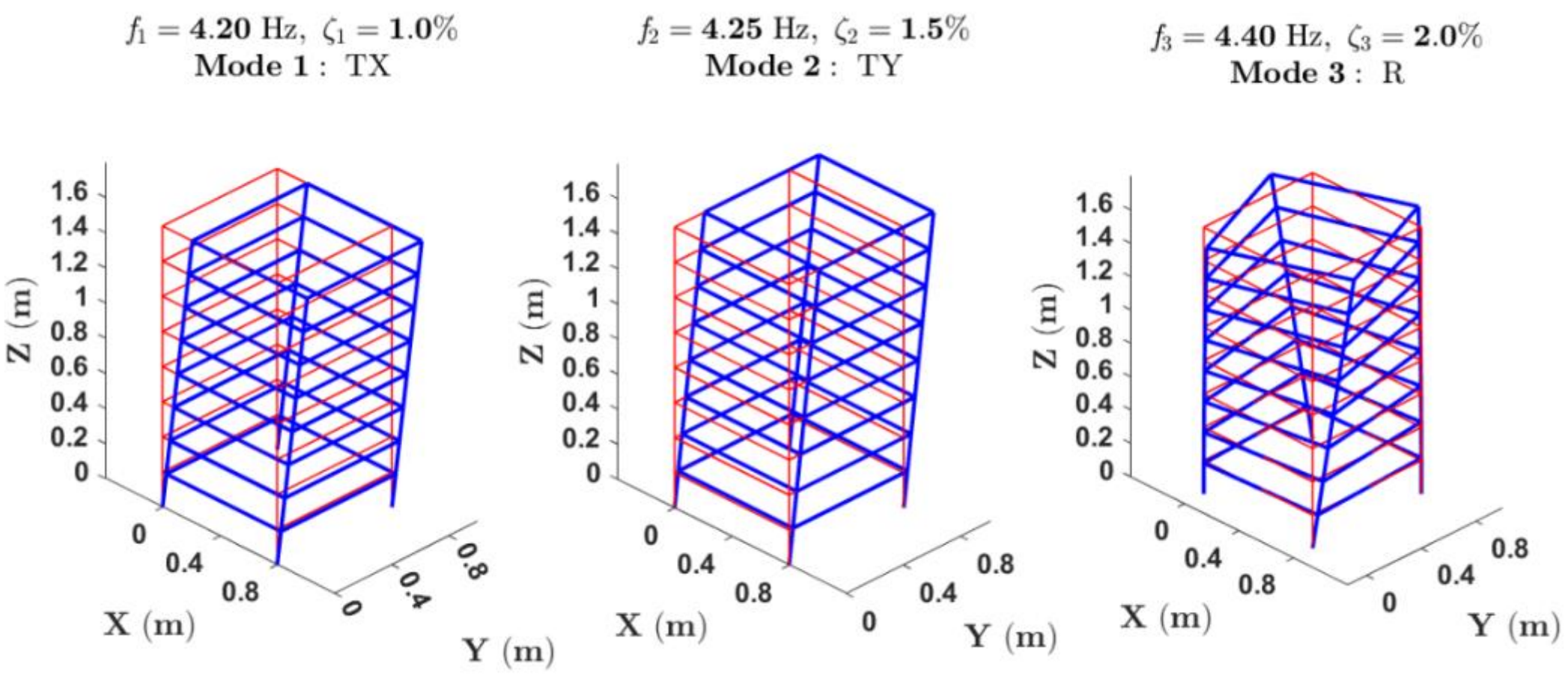

b) True values of modal parameters used for generating data.

**Fig. 1.** Information of the synthetic shear frame model.

The time and frequency domain responses are shown in **Fig. 2**. The first subfigure presents the time history of two representative DoFs over the initial 15 minutes. These time-domain signals reflect the vibrational response of the structure to the assumed modal excitation. The second subfigure displays the auto-PSD, which corresponds to the diagonal elements of the PSD matrix obtained from the time history data. The significant peaks visible in the auto-PSD plot signify resonances in the structure, indicating where potential modes are located. However, these peaks do not explicitly reveal the number of modes that dominate each resonance band. Derived from the eigenvalues of the PSD matrix, the SV spectrum in the third subfigure provides this information. Those bold lines greatly higher than the others correspond to the structural response, while the lower flat lines are attributed to the prediction error. By combining insights from the two spectra, the frequency range for further analysis can be confidently selected, as indicated by the black bracket, and initial estimates of the natural frequencies are provided by the three black circles.

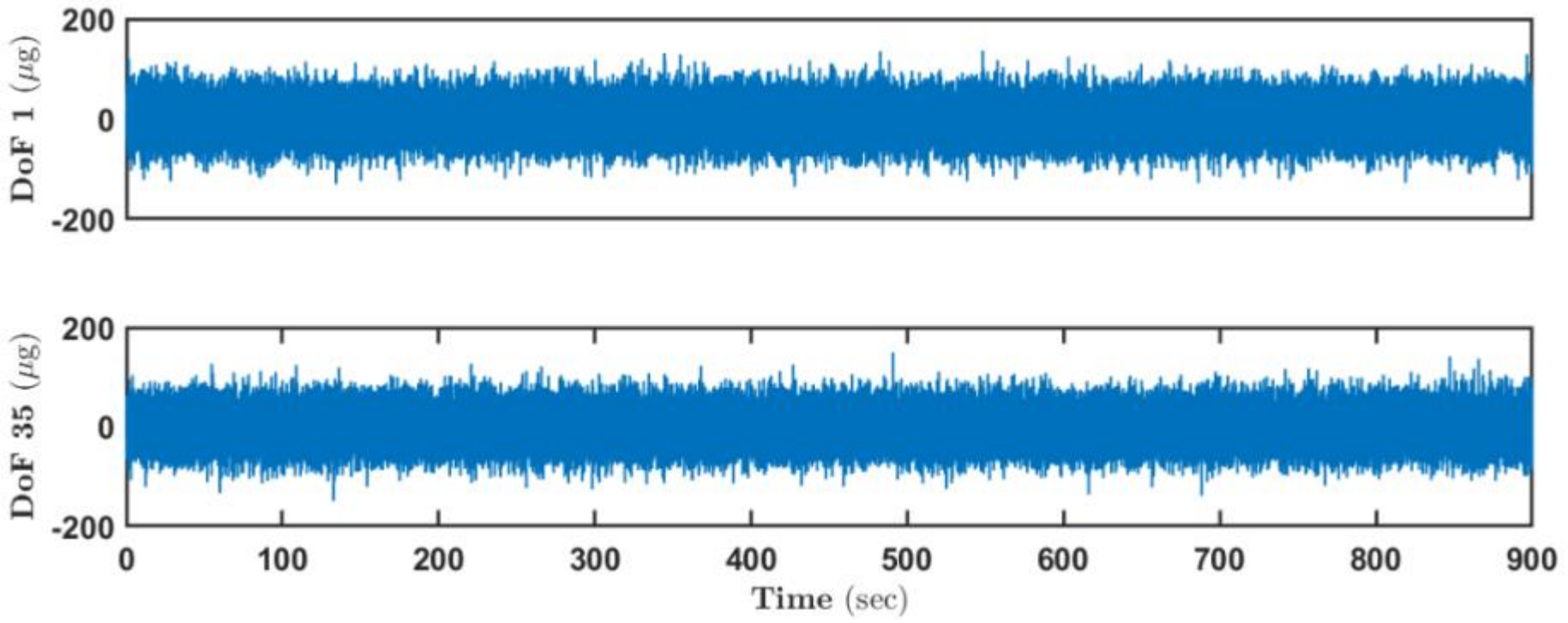

a) Time history.

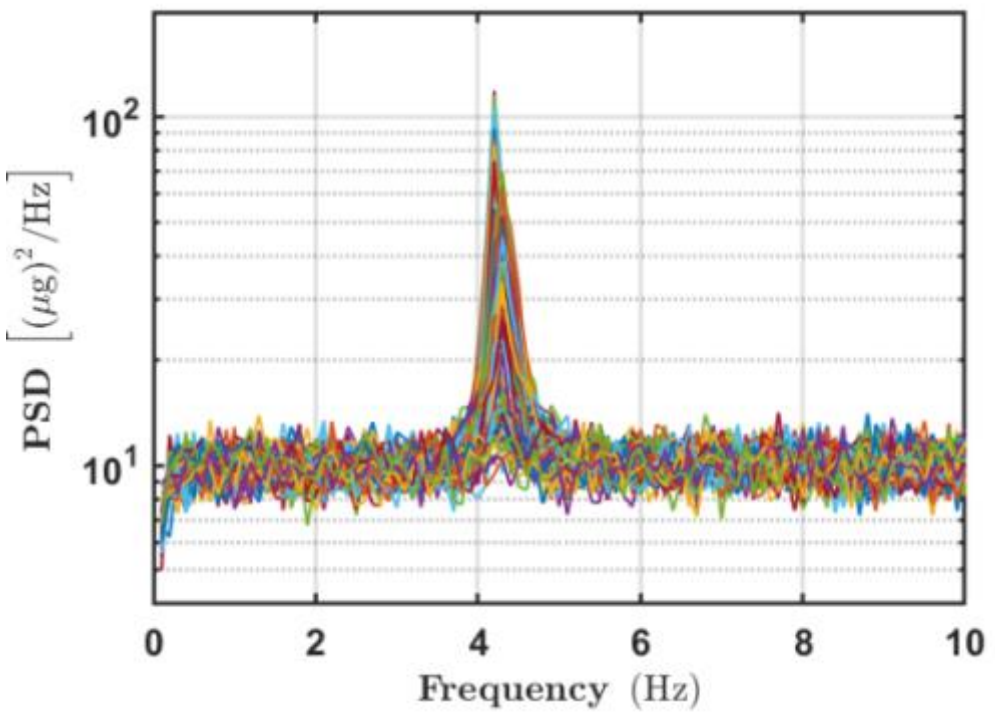

b) Auto power spectrum density.

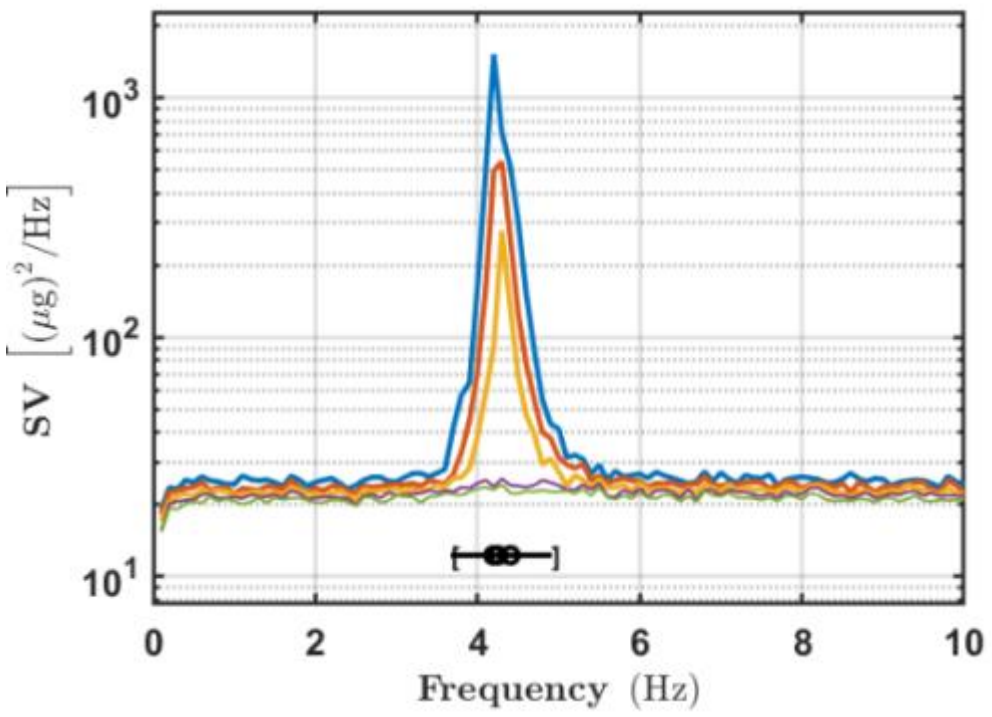

c) Singular value spectrum.

**Fig. 2.** Structural response in time and frequency domains, synthetic example. Only 5 lines for the first five maximum eigenvalues are presented in the SV spectrum to simplify the figure.

## 4.2 Comparative study

To validate the PCM obtained by the proposed method, a comparison with the finite difference method (FDM) and the state-of-the-art (Direct) approach is shown in **Fig. 3**. The results are based on a 4-setup test, with each setup covering two floors from top to bottom. Sensors were fixed at Points (1) and (5) as reference points, while 8 additional sensors were roved to the remaining DoFs. The duration for each setup was 5 minutes to reduce the computational burden of the FDM. The figure illustrates the identified modal parameters of the first mode, along with their associated uncertainties determined by four methods, as shown by those error bars. The modal parameter estimation was performed using the P-EM algorithm [33]. For the uncertainty, FDM1 refers to the case where the Hessian matrix of the NLLF is computed via FDM, and the singularity is handled using a pseudoinverse in Eqn. (14). FDM2 differs from FDM1 in that it applies the null space projection in Eqn. (17) to account for the mode shape norm constraints. The fact that the error bars generated by these two methods are identical confirms the correctness of the null space projection procedure. Next, comparing the error bars from FDM2 and the proposed EM-based method validates the effectiveness of the efficient Hessian matrix assembly method described in Section 3.2. The direct method is also included for reference, and its consistency with the other approaches provides further cross-validation. In all cases, the error bars cover the true values of the modal parameters, indicating that the multi-setup Bayesian FFT method provides an unbiased estimate.

In the multi-setup Bayesian model, modal parameters vary among setups to account for potential time variability. However, the mode shape is considered time-invariant, reflecting its global nature across all DoFs and setups. The identification uncertainty is evaluated as the square root of the sum of the eigenvalues from the mode shape covariance matrix [2]. **Table 3** presents a comparison of the four methods. The Modal Assurance Criteria (MAC) values for all three modes exceed 0.99, indicating high precision in the identified mode shapes. The uncertainty values computed by all four methods also show consistent results. In terms of computational time, the proposed method significantly outperforms the others, being much faster than the direct method, and far more efficient than FDM. Given the high number of DoFs typically required for large-scale engineering structures, this time efficiency becomes even more critical. The FDM, due to its computational and memory demands, is impractical for real-world applications. The direct method also faces significant limitations with large-scale problems. These differences will be further discussed in the next subsection, which analyzes the factors influencing computational time.

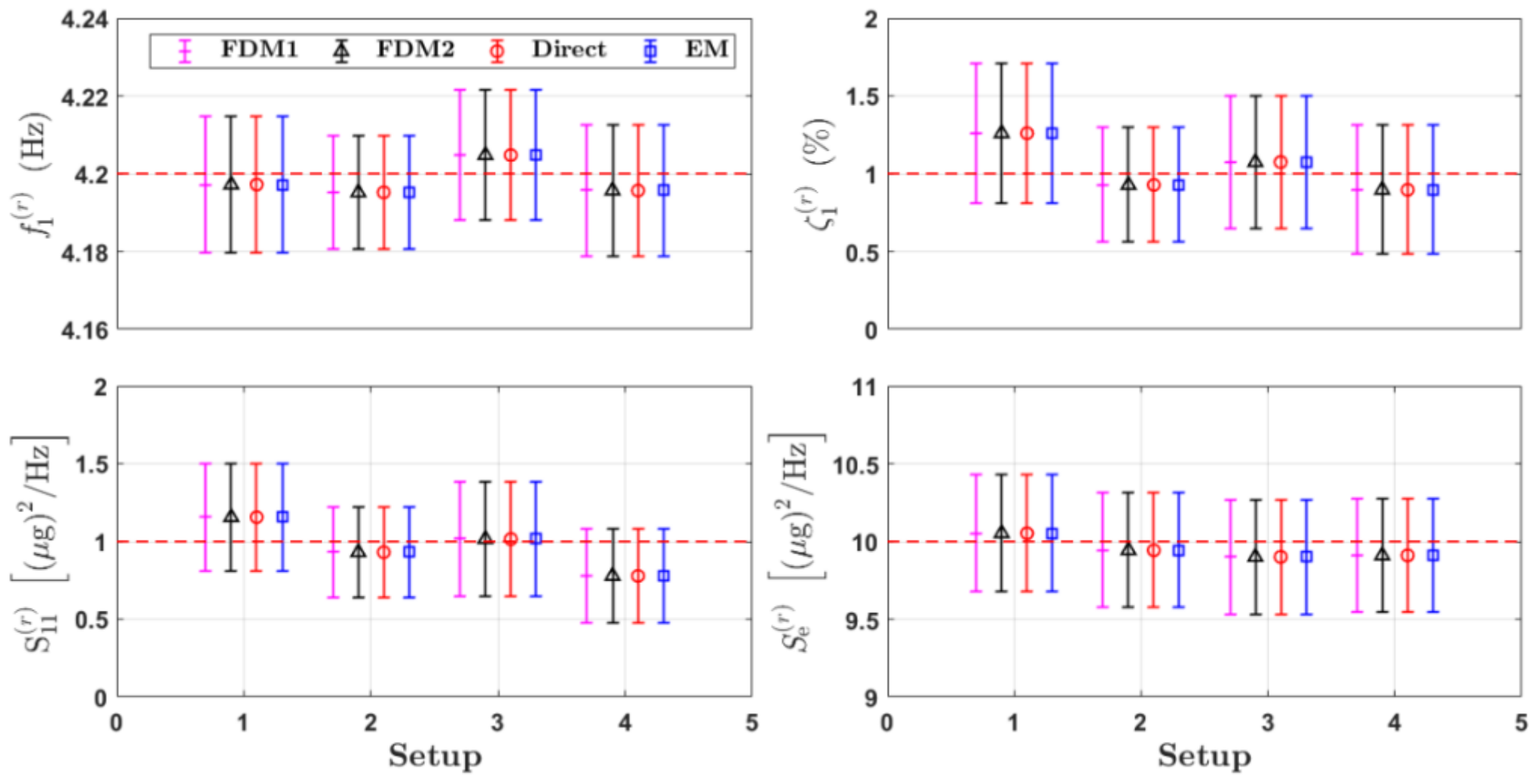

**Fig. 3.** Identification uncertainty comparison for different setups, synthetic data. The red dashed line denotes the true value, and the error bar covers three standard deviations.

**Table 3.** Identification uncertainty comparison of different modes, synthetic data. The MAC is computed between the identified mode shape by the P-EM method and the true one.

| Mode | MAC | c.o.v. [%] | | | |
|---|---|---|---|---|---|
| | | FDM1 | FDM2 | Direct | EM |
| TX | 0.9964 | 8.67 | 8.67 | 8.67 | 8.67 |
| TY | 0.9961 | 11.92 | 11.92 | 11.95 | 11.95 |
| R | 0.9966 | 10.55 | 10.55 | 10.60 | 10.60 |
| Computational time (sec) | | 1441 | 1438 | 4.18 | 0.88 |

## 4.3 Efficiency analysis

The main factor affecting computational time in the uncertainty quantification is the Hessian matrix assembly. The proposed method leverages partial decoupling of parameters in the complete-data NLLF, reducing the need to compute and store numerous intermediate terms, as required by the direct method. This results in lower time and memory consumption compared to the direct approach.

The computational time for assembling the Hessian matrix depends on two key factors: the number of parameters, which increases with the number of setups, DoFs, and modes, and the number of FFT points, which is determined by the test duration when the band is selected.

In practice, the number of modes within a frequency band is typically no more than three, providing limited variability. Besides, the number of measured DoFs usually increases with the number of setups, which allows us to analyze the impact on computational time. In this test, suppose 6 sensors were used: 2 fixed at Points (1) and (5) as references, while the remaining sensors roved across the floor from top to bottom. The number of setups ranged from 2 to 8 until all 34 points were covered by the last case. As a result, the number of parameters linearly increased from 92 for 2 setups, to 332 for 8 setups. The duration for each setup was set as 30 minutes. The consistency of mode shape uncertainties between all three methods is shown in **Fig. 4**, validating that the overall Hessian matrix remains equivalent in the two methods. The computational time for each scenario is highlighted in **Fig. 5**, where the proposed method outperforms the direct approach by at least six times. The computational times shown in the figure represent the mean of ten repetitions, all of which were run on the same desktop equipped with an Intel® Core™ i7-10700K processor running at 3.80 GHz and 16.0 GB of RAM. The error bars display five standard deviations from these repetitions. Despite the small fluctuations, the overall trend remains stable across the runs, indicating consistent performance even in single executions. This efficiency stems from two factors. First, the partial decoupling of parameters in the complete-data NLLF, combined with the matrix calculus, optimizes compatibility with matrix-based operations in programming environments like MATLAB. Second, the sparse Hessian matrix assembly transforms large-scale matrix calculations into low-dimensional ones, significantly improving computational speed.

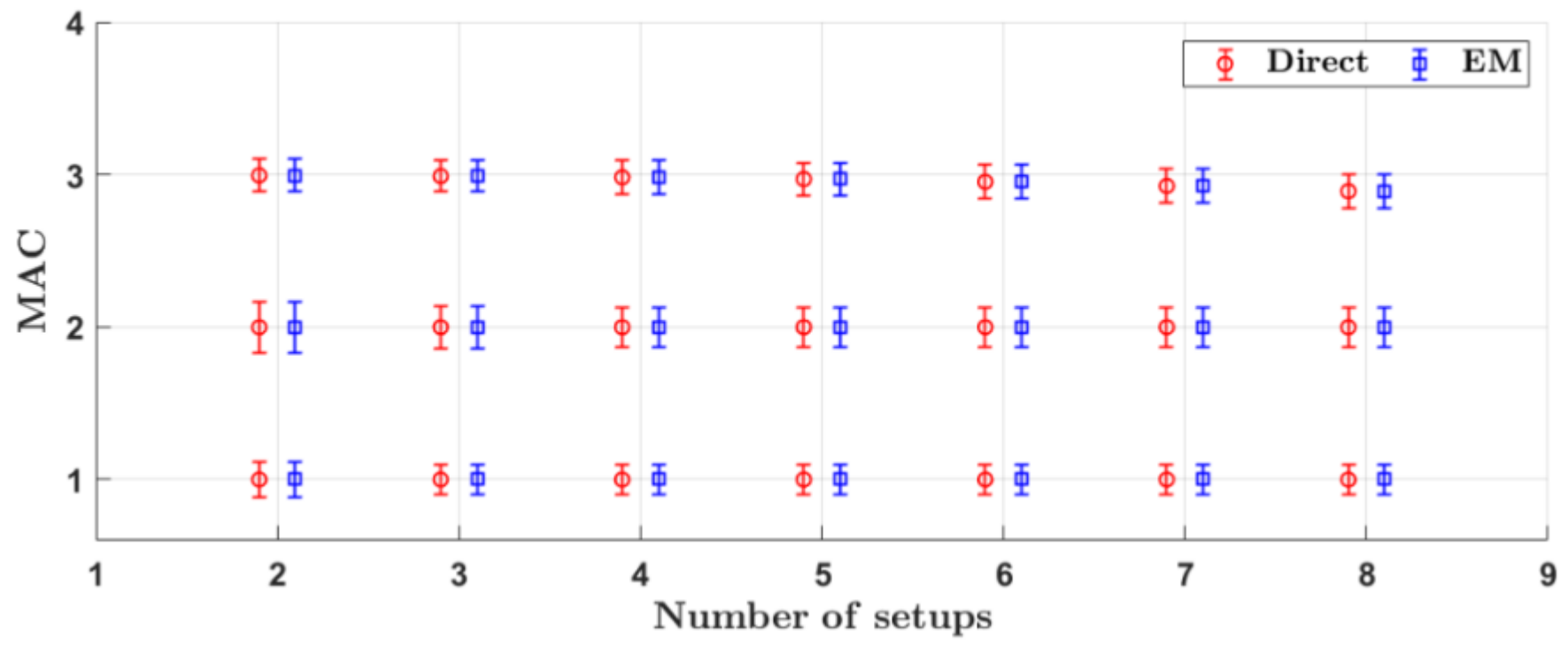

**Fig. 4.** Mode shape MAC comparison with different numbers of setups. The MAC is calculated between the identified mode shape and the corresponding true one and it has been translated upward by 1 and 2 for the second and third modes. The error bar covers three standard deviations.

To analyze the effect of test duration on uncertainty and computational time, we conducted an 8-setup test, similar to the configuration described earlier. The test covered all 34 points using 6 sensors, with 2 fixed at Points (1) and (5) as references and the remaining 4 sensors roving across the structure. The duration of each setup ranged from 15 to 30 minutes, resulting in a linear increase in the number of FFT points, from 1123 to 2245. The comparison of mode shape uncertainty between the two methods is presented in **Fig 6**. Again, the high consistency with the same length of error bar is observed. Moreover, as the test duration increases, the error bars shrink, indicating a reduction in uncertainty. This matches the fact that more FFT points provide more information for analysis, leading to improved accuracy. The computational times for the two methods are shown in **Fig 7**. As expected, the consuming time increases as the number of FFT points increases for both methods. However, the increase is more pronounced for the direct method. Specifically, the computational time for the direct method ranges from 20 seconds for the 5-minute duration to over 36 seconds for the 15-minute duration. In contrast, the proposed method shows a much smaller increase, ranging from around 3 seconds to 6 seconds. This efficiency is attributed to the proposed sparse Hessian assembling strategy, which minimizes unnecessary computations. Unlike the direct method, the proposed method is less sensitive to the increase in FFT points, making it more efficient in handling longer test durations.

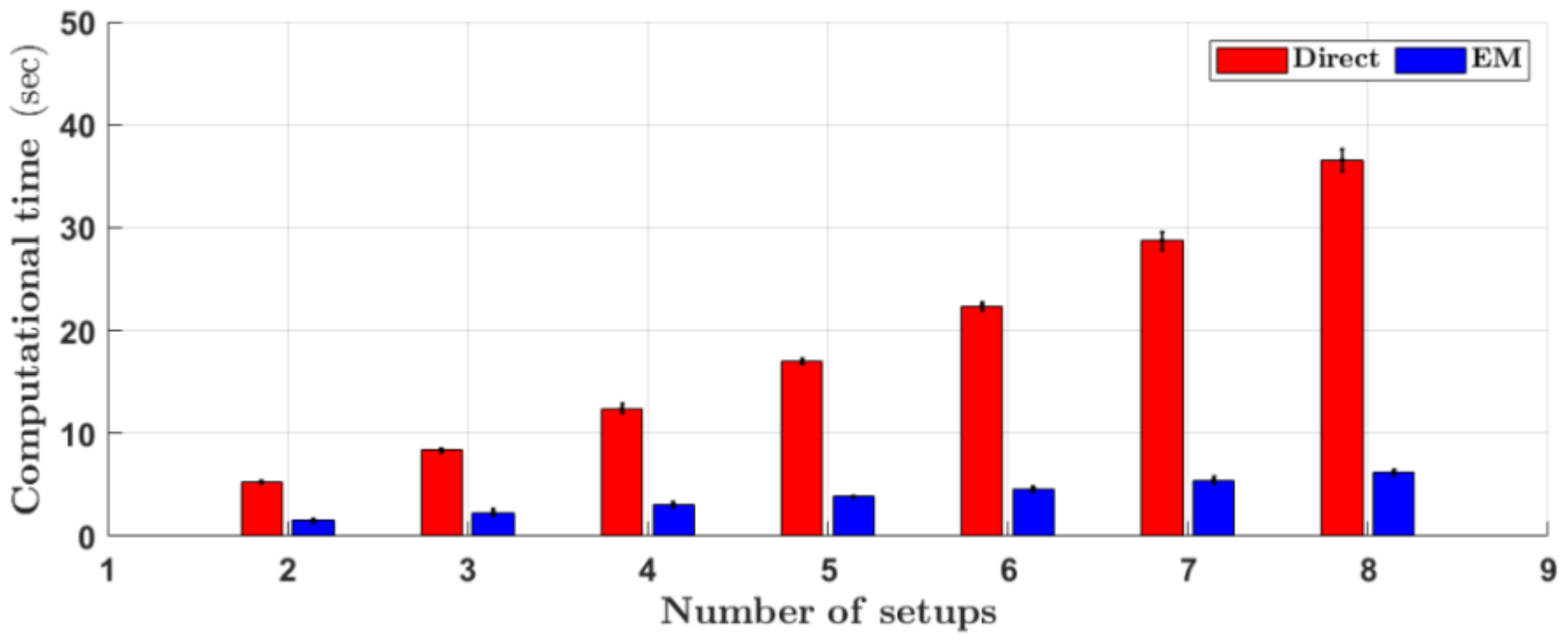

**Fig. 5.** Computational time comparison with different numbers of setups. The error bar above the histogram covers five standard deviations coming from ten repetitions.

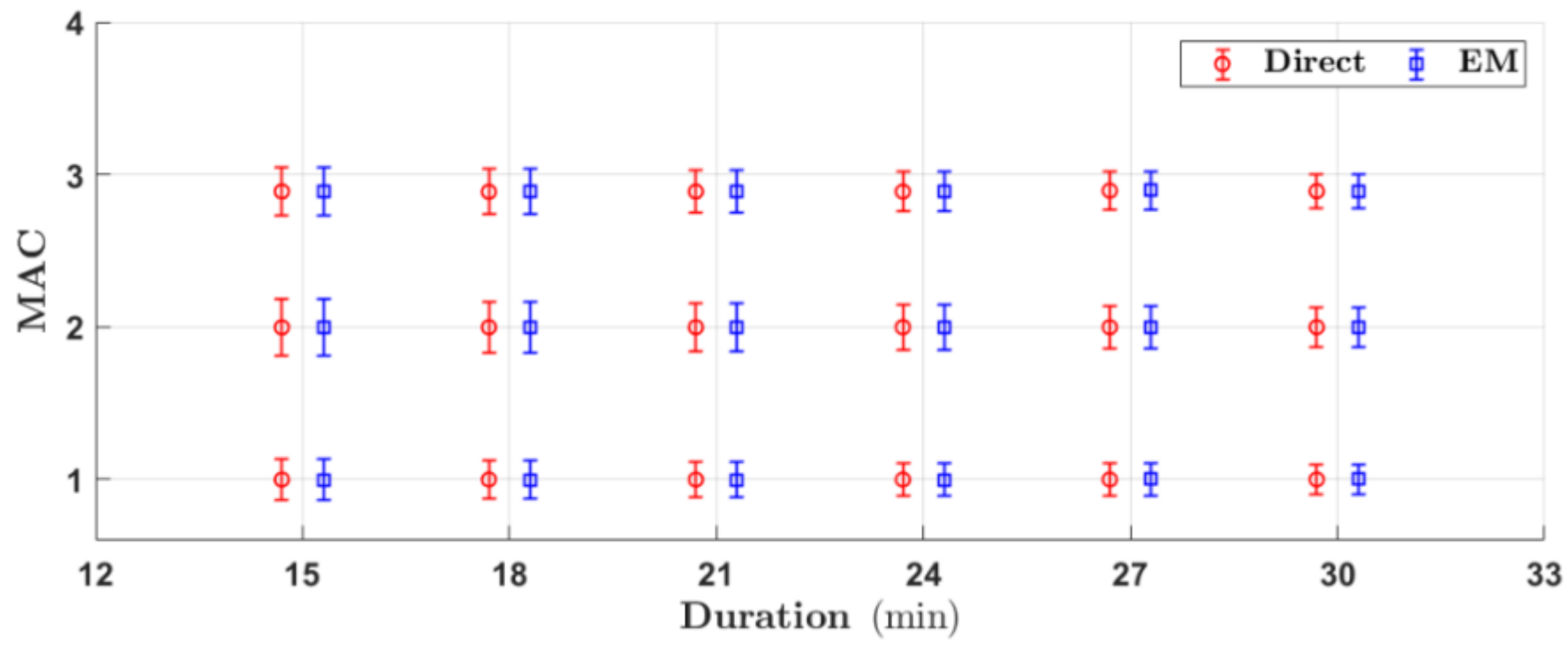

**Fig 6.** Mode shape MAC comparison with different durations. The MAC is calculated between the identified mode shape and the corresponding true one and it has been translated upward by 1 and 2 for the second and third modes. The error bar covers two standard deviations.

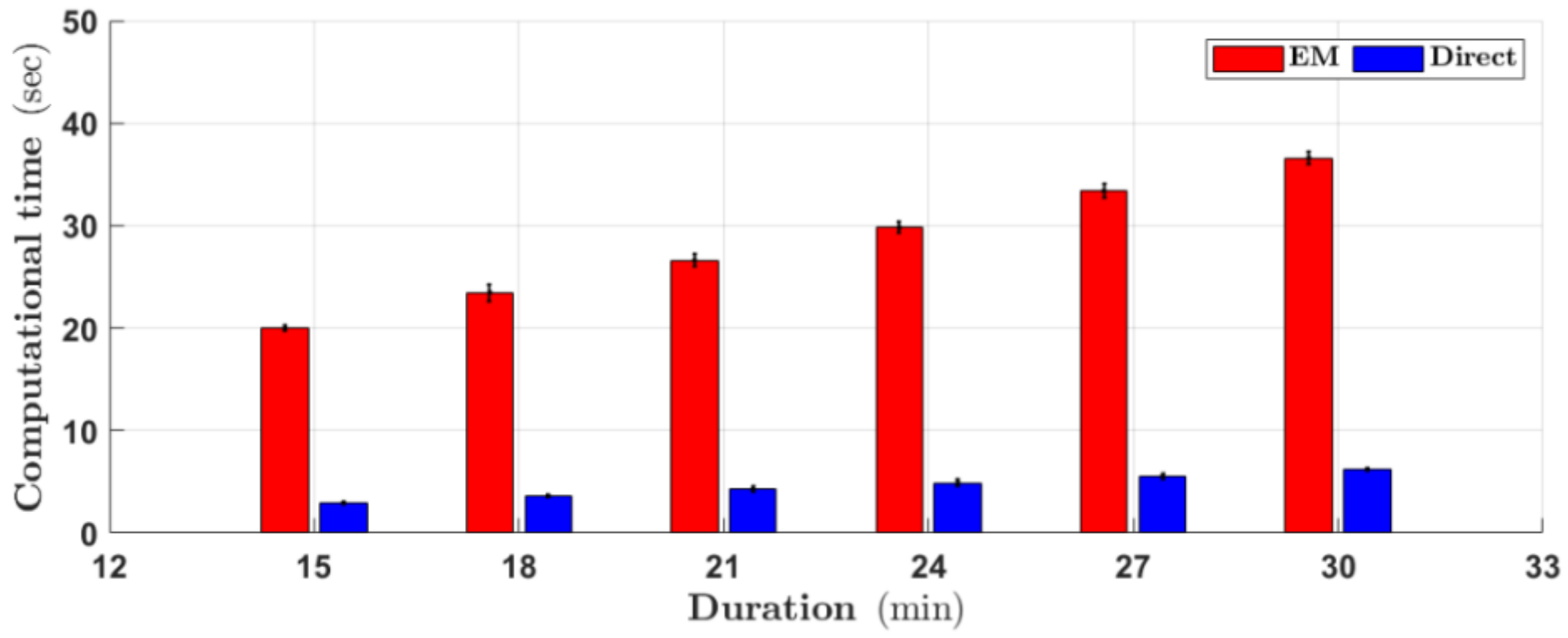

**Fig 7.** Computational time comparison with different durations. The error bar above the histogram covers three standard deviations coming from ten repetitions.

# 5 Field test application: YR-ICC Building

In this section, we assess a real-world engineering structure to evaluate the feasibility and performance of the proposed method. The structure under study is the Yangtze River International Conference Center (YR-ICC) located in Nanjing, China (**Fig. 8** a)). This multi-purpose complex primarily hosts international conferences and business events. The building dimensions are 81 m in length, 15 m in width, and 93 m in height. The significant difference between the two horizontal axes leads to a variation in physical properties, such as stiffness, in each direction, influencing modal behaviors. For such a high-rise building, a conventional single-setup vibration test cannot provide sufficient spatial resolution to capture the mode shapes of the building accurately. Thus, a multi-setup ambient vibration test was conducted before the building was officially opened to the public.

## 5.1 Test configuration

As illustrated in **Fig. 8** b), the testing process involved 14 setups using 8 Kinemetrics Etna 2 accelerometers, each measuring 3 channels, providing a total of 204 measured DoFs. A standalone instrument setup is shown in **Fig. 8** c), which includes a force-balanced accelerometer, an external battery, and a GPS module. The accelerometer is equipped with a built-in Linux system that integrates data acquisition and storage. The power supply is provided by a lead-acid battery, which requires a full 24-hour charge to reach 14.8 V. The GPS module is employed to synchronize different measurement channels. Before the test, the 8 sensors were calibrated using a huddle test [44], as depicted in **Fig. 8** d). The noise level for each instrument is in the order of 0.1 $\mu\mathrm{g}/\sqrt{\mathrm{Hz}}$ in the frequency range of 0.1 to 40 Hz. Three typical configurations were conducted during the test. In the first setup, sensors were positioned on the top floor, 93 meters above ground. The three points marked in red served as reference points, while the remaining 5 sensors roved from the top floor to a height of 21 meters, completing the next 8 setups. Before starting Setup 10, the external battery of a sensor unexpectedly failed. Consequently, only 4 rover sensors were available for the final 5 setups. Despite this setback, the multi-setup test demonstrated flexibility and robustness in real-world conditions. However, certain DoFs remained unmeasured due to safety concerns or inaccessibility. For example, only one sensor could be placed on one side along the X direction in most cases, reflecting the unavoidable variability in field test configurations.

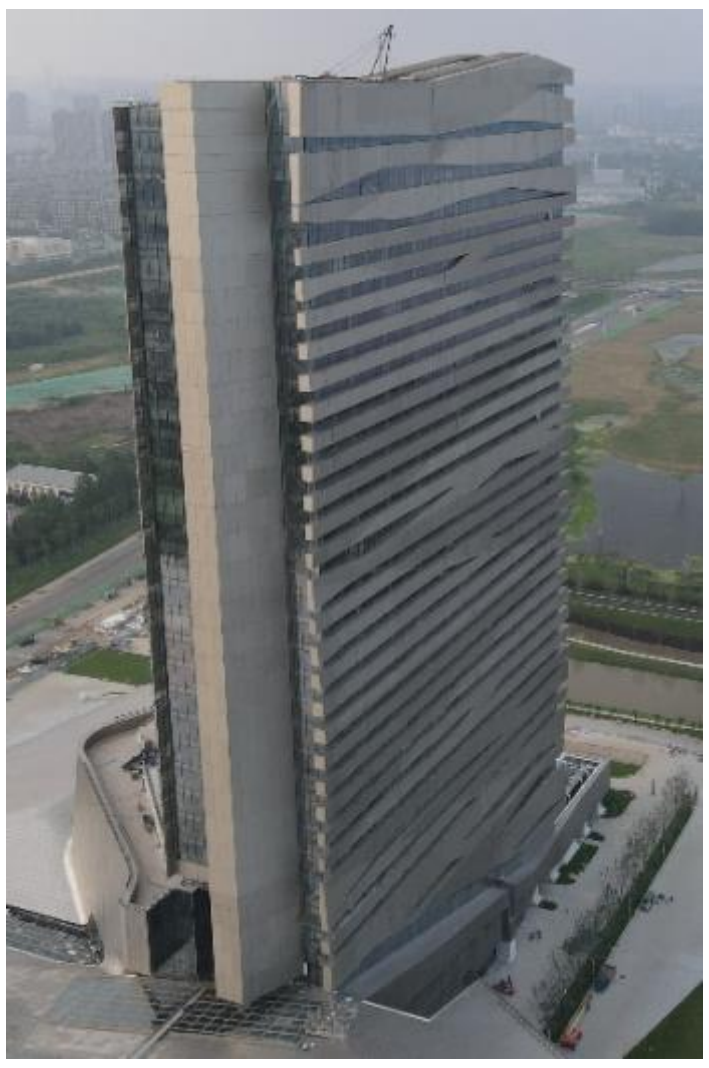

a) Field photo.

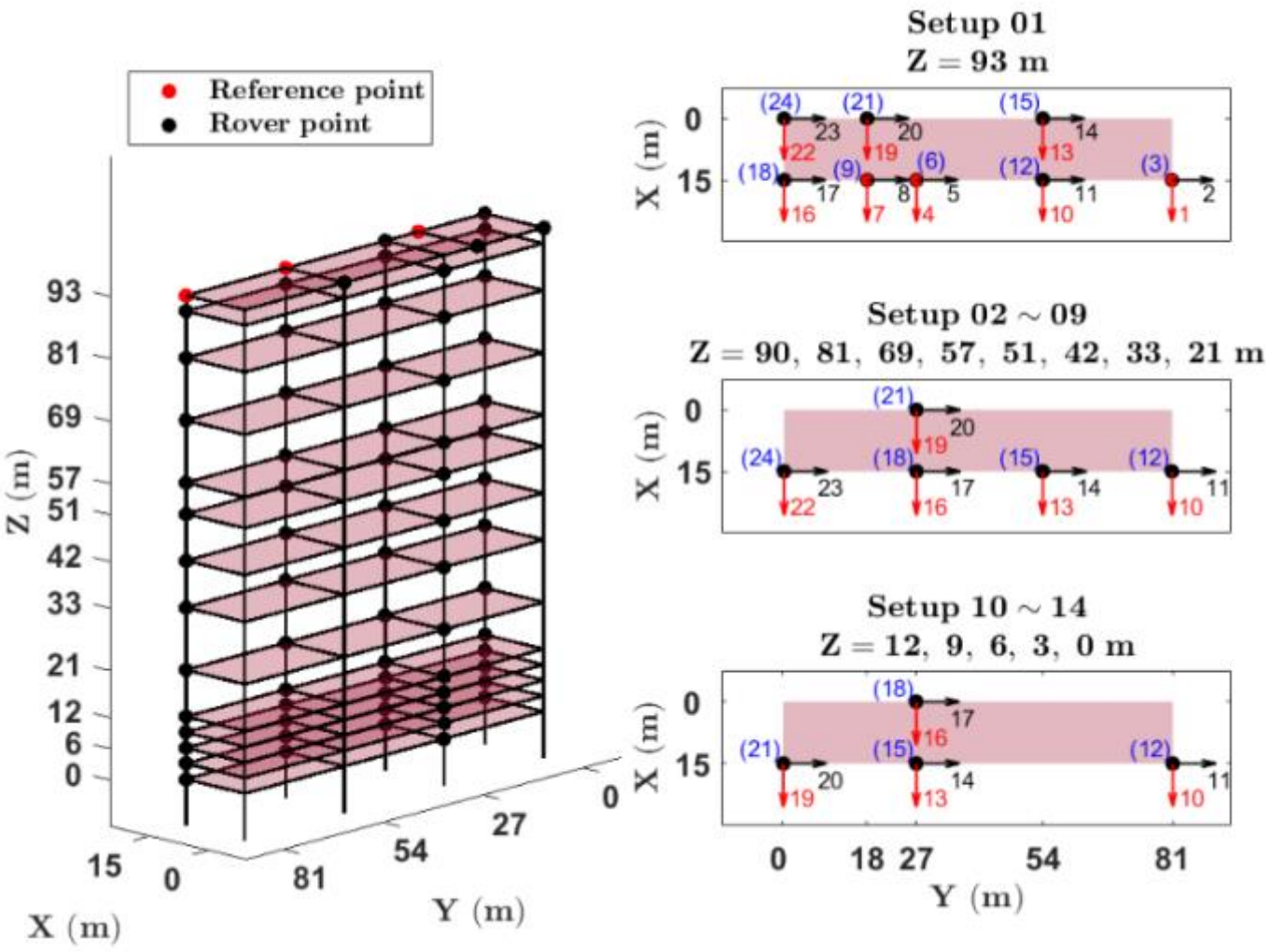

b) Test configuration. The blue numbers in the parenthesis indicate the DoFs in the Z-axis direction.

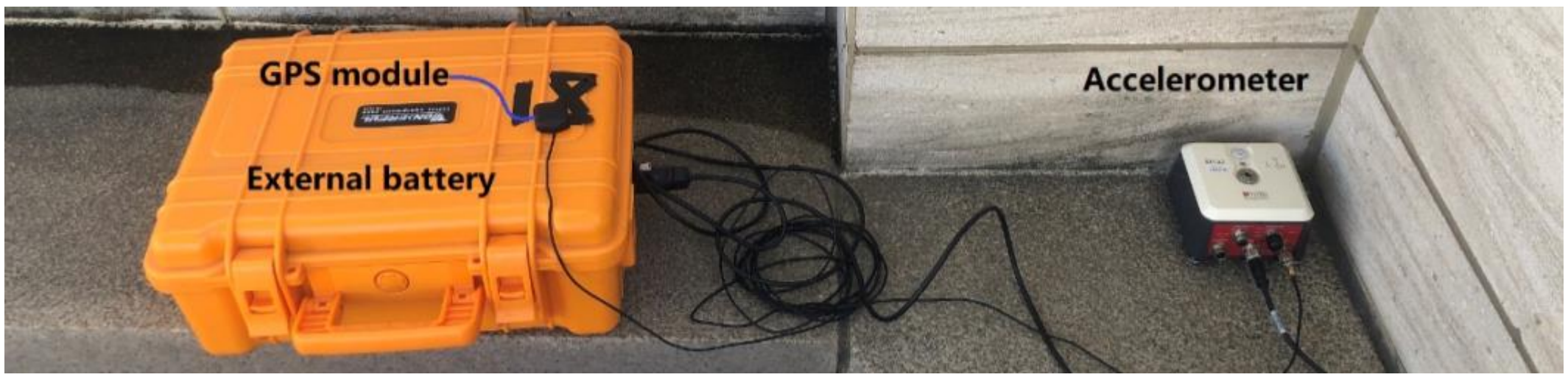

c) Single instrument setup.

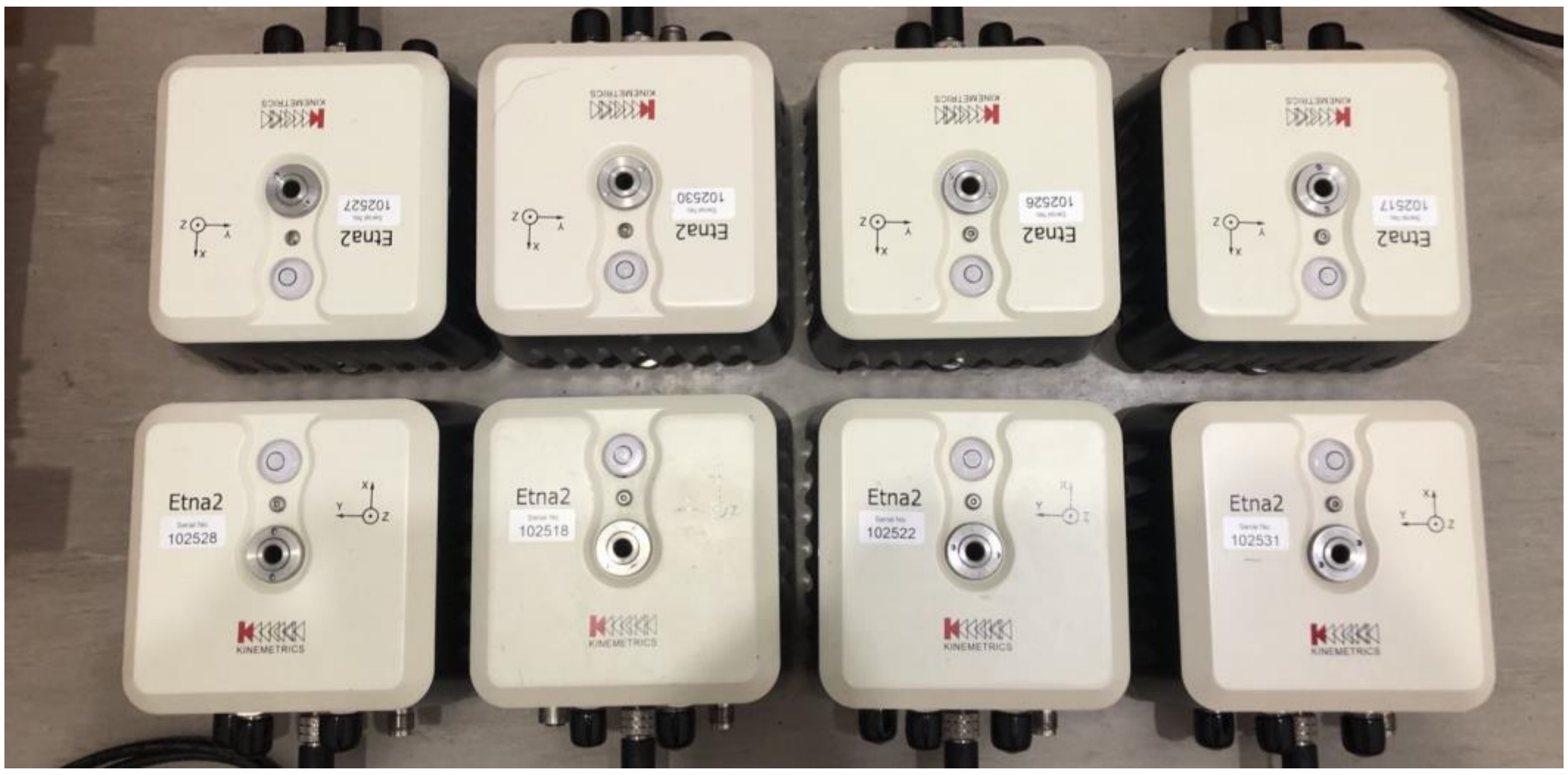

d) Multi-Sensor calibration.

**Fig. 8.** Information of the YR-ICC Building.

Ambient response data was collected for 10 minutes in each setup at a sampling rate of 100 Hz, with additional time required for moving equipment between setups. In total, the testing process took approximately 4 hours. A typical time history of 10 min for the first two DoFs from the initial setup is shown in **Fig. 9** a). The corresponding frequency domain representations are shown in the following two subfigures. Higher modes exhibit a lower SNR, as indicated by the small discrepancies between the bold lines (dominant modes) and the remaining lines (noise and less significant modes). This is mainly because ambient excitations, such as wind, are typically dominated by low-frequency components. Additionally, the lower singular values around the higher modes receive contributions from the dynamics of lower modes, further elevating the 'noise' level. Six frequency bands were selected to cover the possible resonance ranges. Of these, two bands contain three closely-spaced modes, one band contains two closely-spaced modes, and the remaining three bands contain well-separated modes.

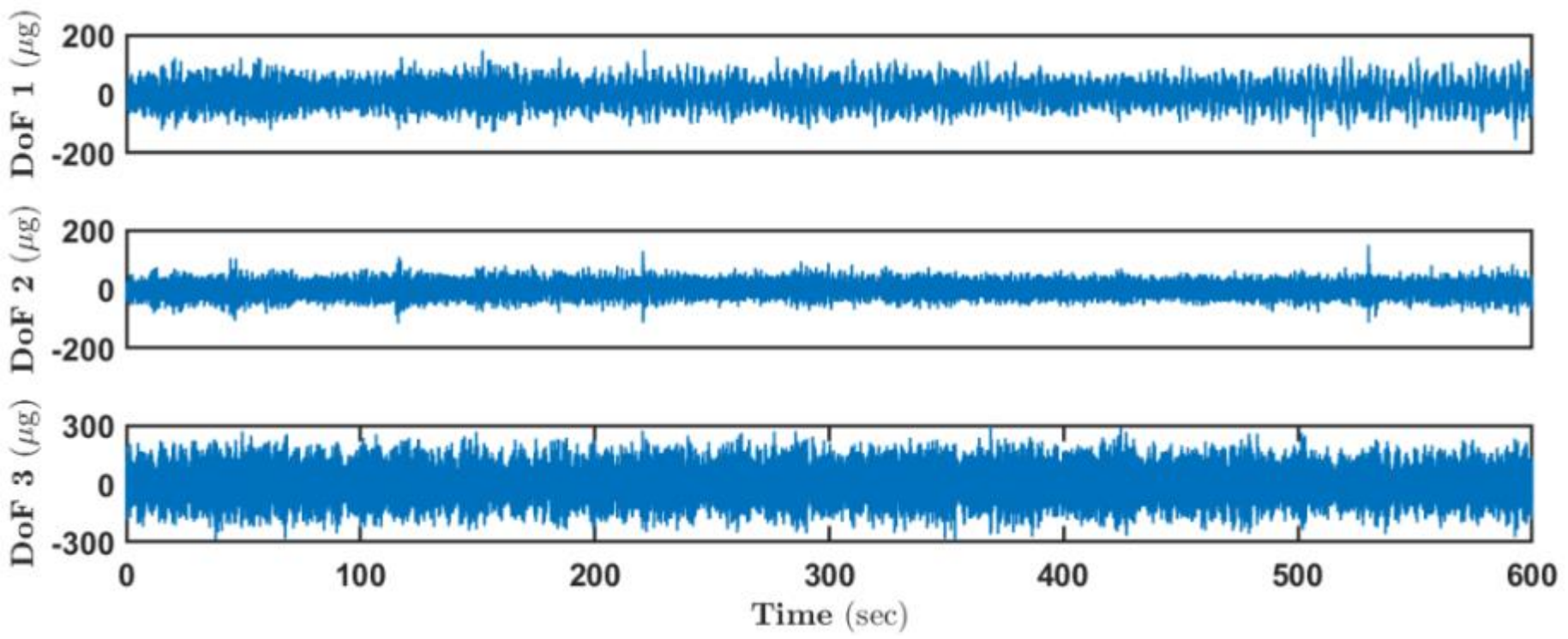

a) Time history.

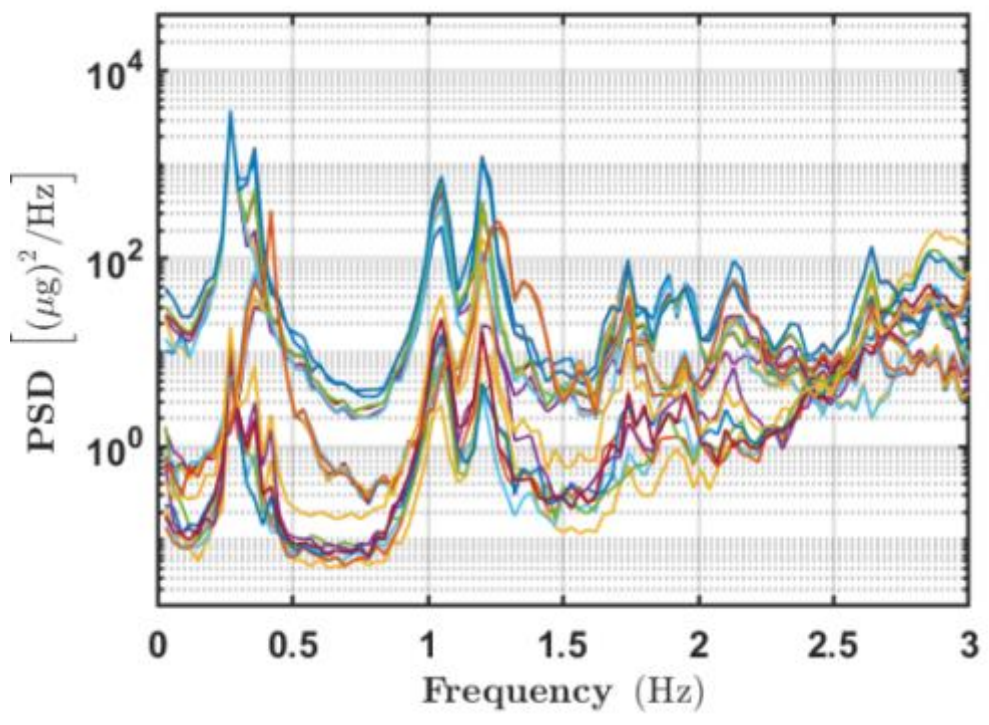

b) Auto power spectrum density.

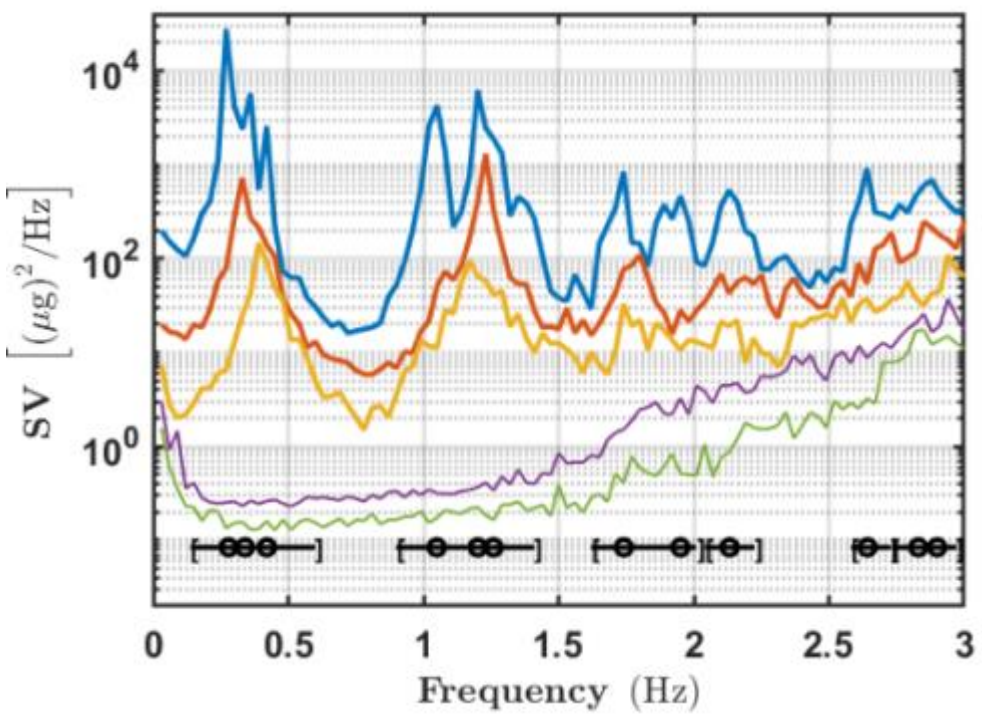

c) Singular value spectrum.

**Fig. 9.** Structural response in time and frequency domains, YR-ICC Building.

## 5.2 Identification result

Using the P-EM algorithm [33], a total of twelve modes were identified, as detailed in **Fig. 10**. Since the natural frequency and damping ratio vary across setups, their average values are used to represent each mode. The fundamental frequency of the structure is approximately 0.277 Hz, while the twelfth mode remains below 3 Hz, indicating the flexibility of the structure as a high-rise building. The damping ratios for the first eleven modes are below 3%, consistent with typical ambient conditions. However, the last mode in the Z direction exhibited a significantly higher damping ratio, exceeding 6%. This is unusual under ambient excitation, suggesting a unique mechanism. As shown in **Fig. 9** a), vertical vibrations (primarily excited by ground motion) are notably stronger than wind-induced translational vibrations. Due to the amplitude-dependent nature of damping, this higher value is reasonable and highlights the effect of ground-induced vertical excitation on the structural response.

An interpolation method [45] was applied to predict the mode shape values for the unmeasured DoFs in **Fig. 10**. This method is based on the rigid floor hypothesis, which assumes that the floor deforms rigidly without bending. The deformed motion is fully described by a rotation matrix and a translation vector. Dividing each 15-by-81-meter floor into three rigid 15-by-27-meter sections, a singular value decomposition-based least-squares approach was used to determine the rotation matrix and the translation vector from the measured DoFs. They were further adopted to estimate the mode shape values for the unmeasured DoFs, allowing for a complete representation as shown in the figure. From the spatial distribution of the modes, four translational modes in two directions, one vertical mode, and three torsional modes were identified. Each mode is labeled according to its vibrational nature. For example, TX1 refers to the first translational mode in the X direction, while R1 denotes the first rotational mode. The percentage shown after each mode name reflects the uncertainty of the mode shape, calculated by the proposed method. These uncertainties are all within a few percent, demonstrating a high level of precision in the modal identification.

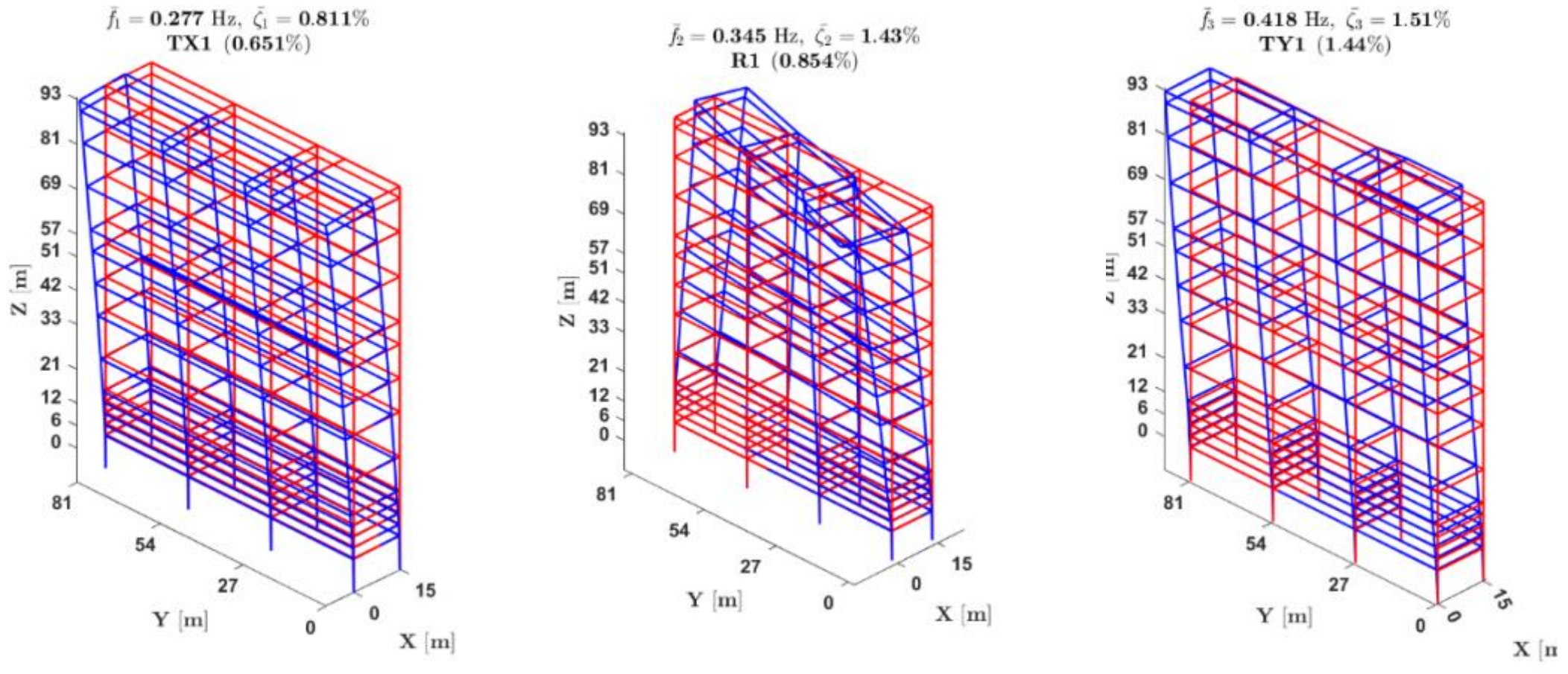

a) Mode 1. b) Mode 2. c) Mode 3.

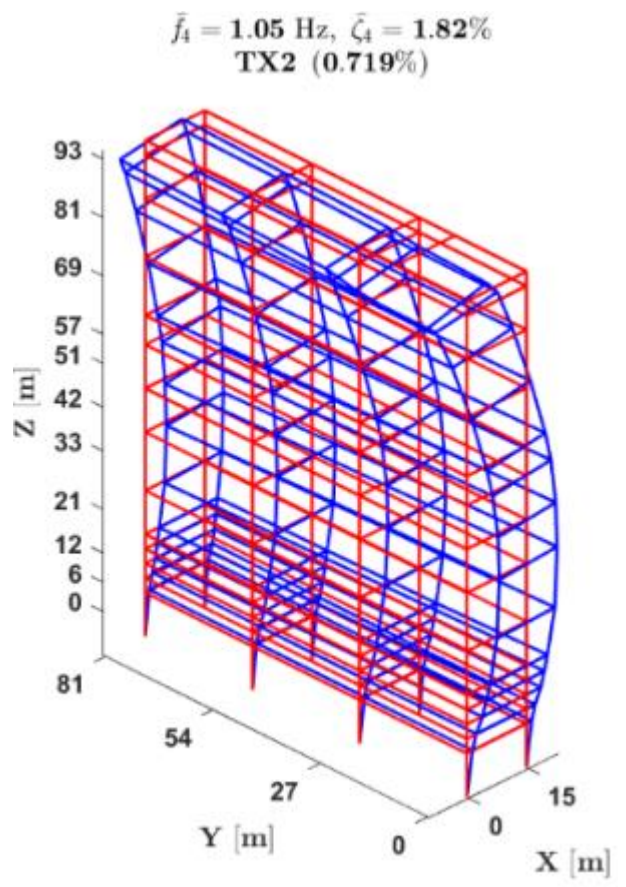

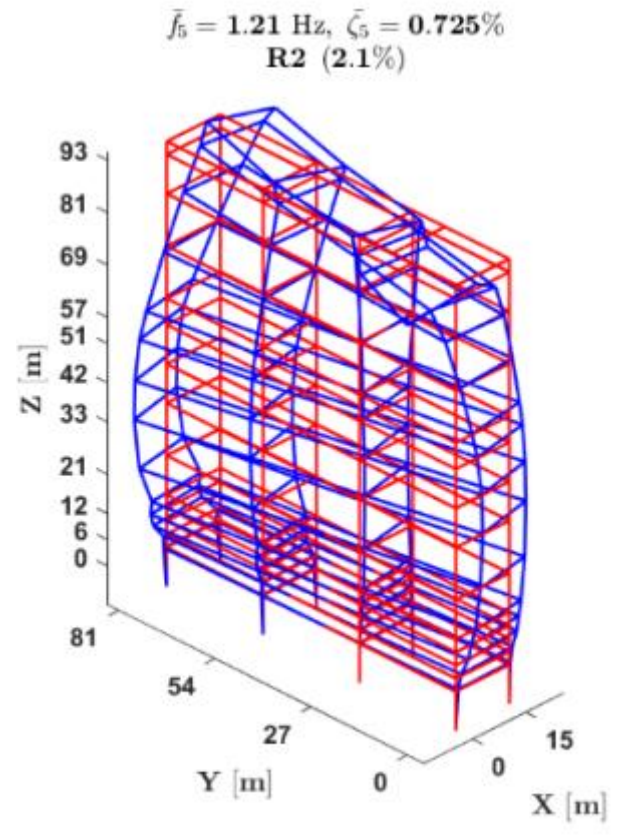

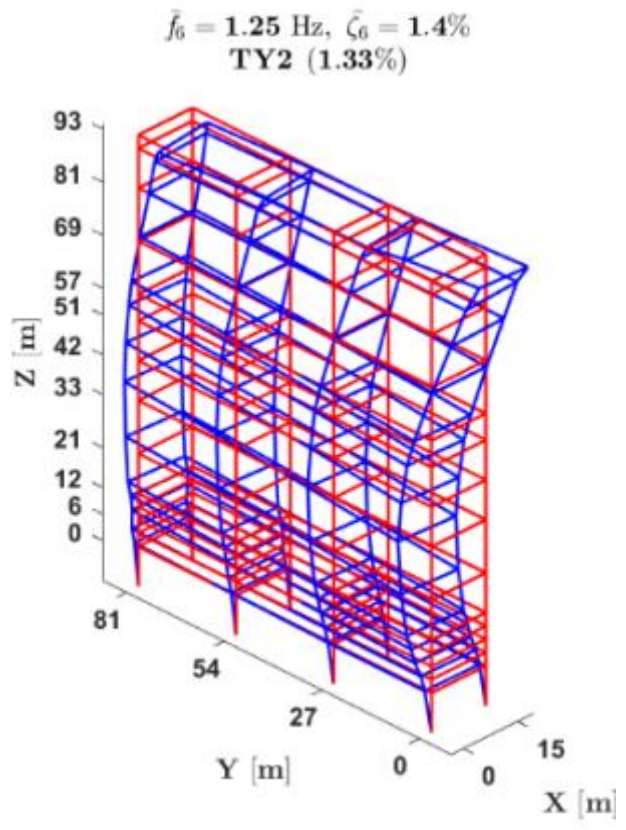

d) Mode 4. e) Mode 5. f) Mode 6.

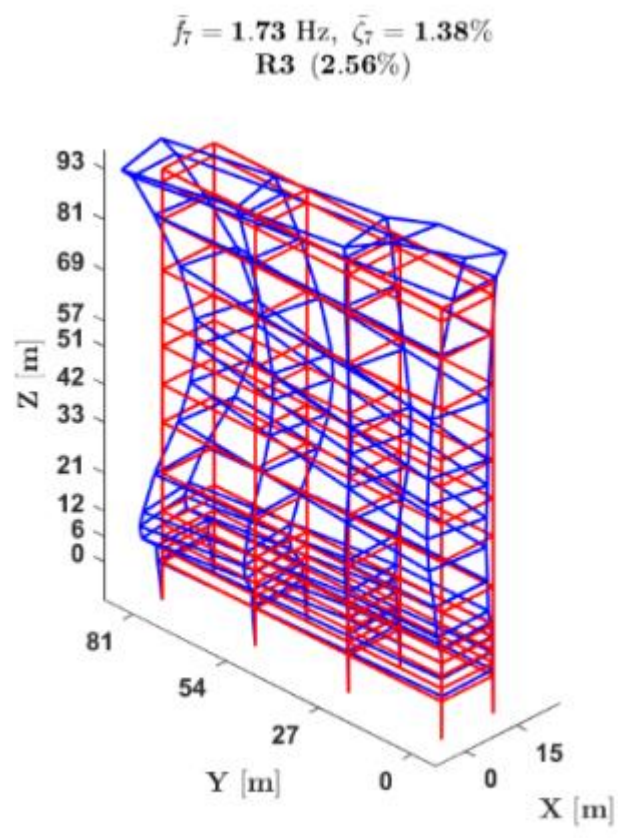

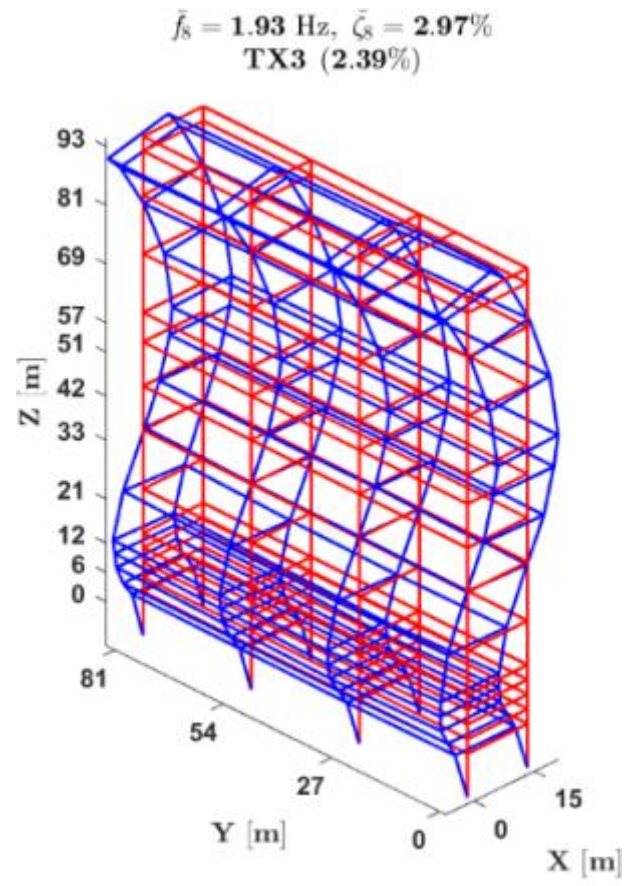

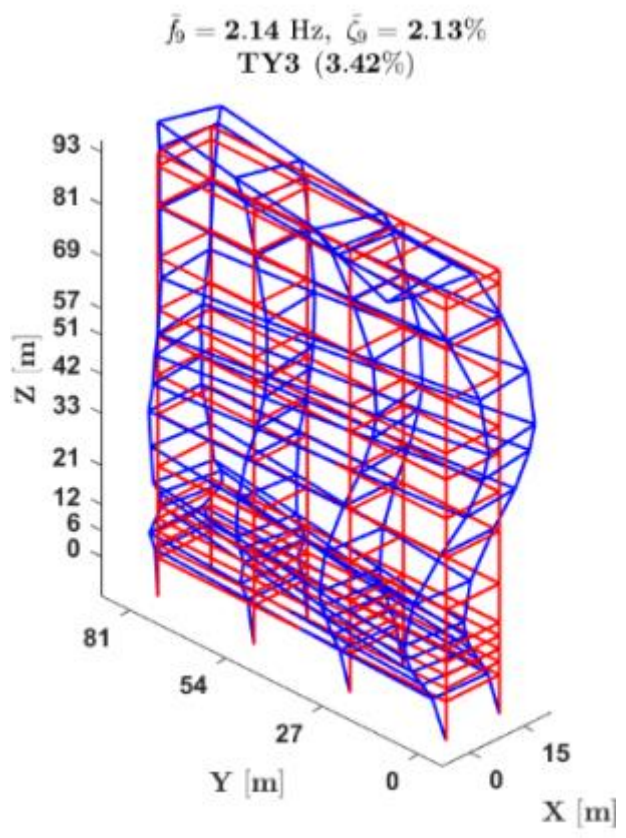

g) Mode 7. h) Mode 8. i) Mode 9.

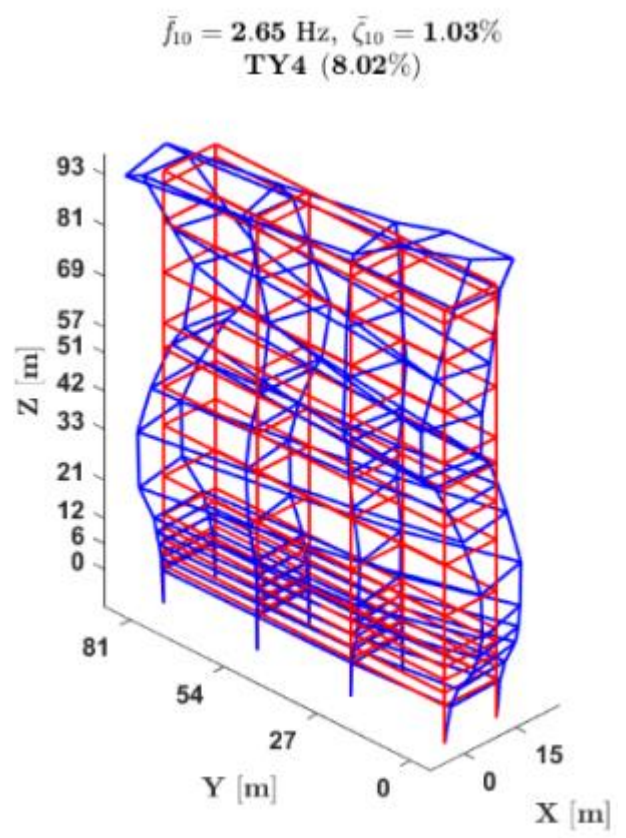

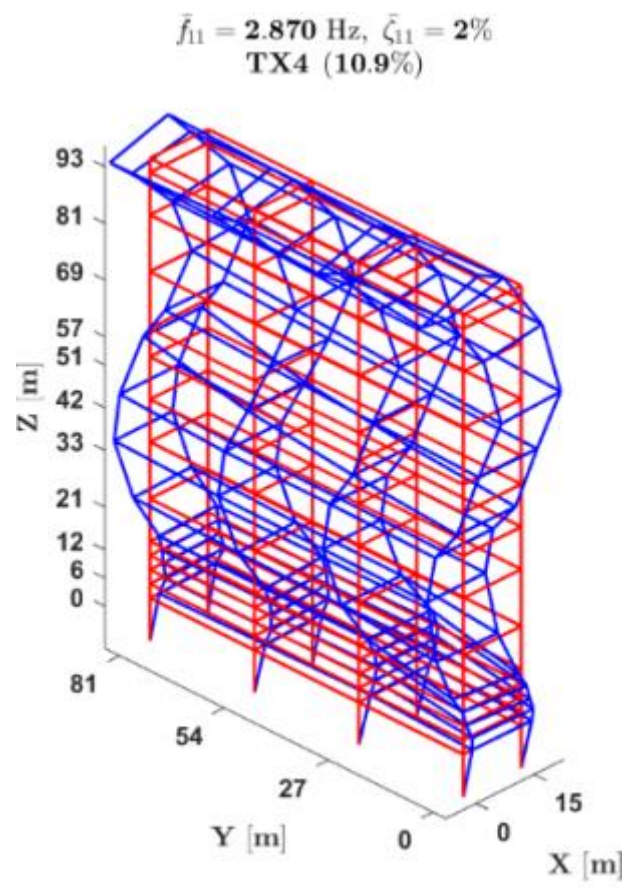

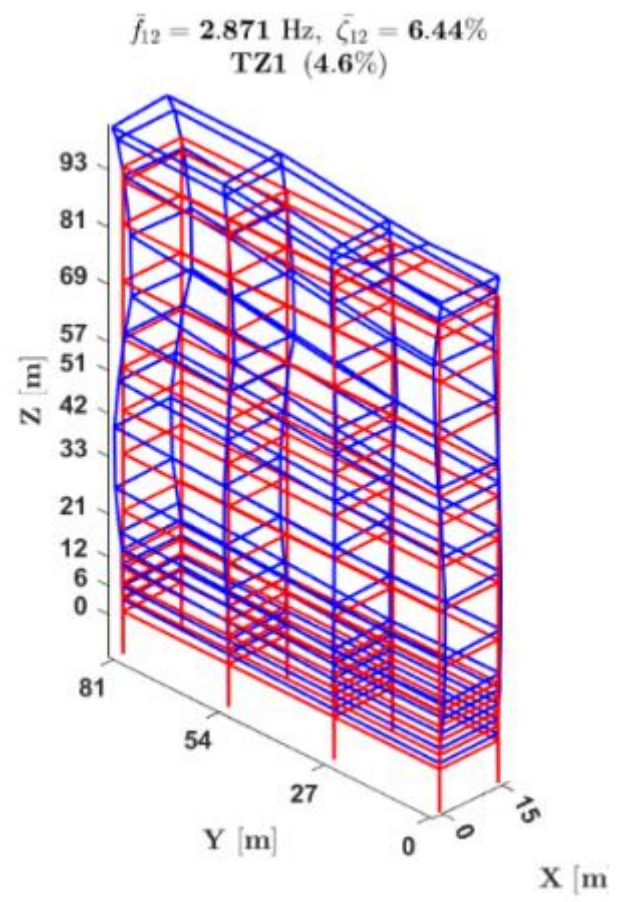

j) Mode 10. k) Mode 11. l) Mode 12.

**Fig. 10.** Identified mode shapes, YR-ICC Building. The red lines denote the original state while the blue lines denote the deformed one; the number in the bracket denotes the posterior c. o. v. of the mode shape.

A comprehensive comparison of the uncertainties provided by the direct and proposed methods is presented in **Fig. 11**. For each band, a representative mode was selected, and both methods yielded similar results, as illustrated by the alignment with the 1:1 dotted line. In terms of mode shape uncertainty, all modes demonstrate consistent results between the two methods. The computational efficiency is shown in **Fig. 12**. Using the same computer as in the previous example, each method was executed only once, as the substantial performance differences rendered results unaffected by minor run-to-run variations. For the first two challenging bands, which contain three closely-spaced modes, the direct method required over 30 seconds to produce the same results that the proposed method achieved in just 3 seconds. This stark contrast highlights the practical need for the proposed fast method. In the third and last bands, which contain two closely-spaced modes, the direct method took 10.79 and 7.17 seconds, respectively, compared to just 0.88 and 0.49 seconds for the proposed method, still showing a significant difference. For the remaining well-separated modes, the direct method took several seconds, which remains practical. However, the proposed method consistently required about 0.1 seconds, making it a superior choice in terms of computational efficiency.

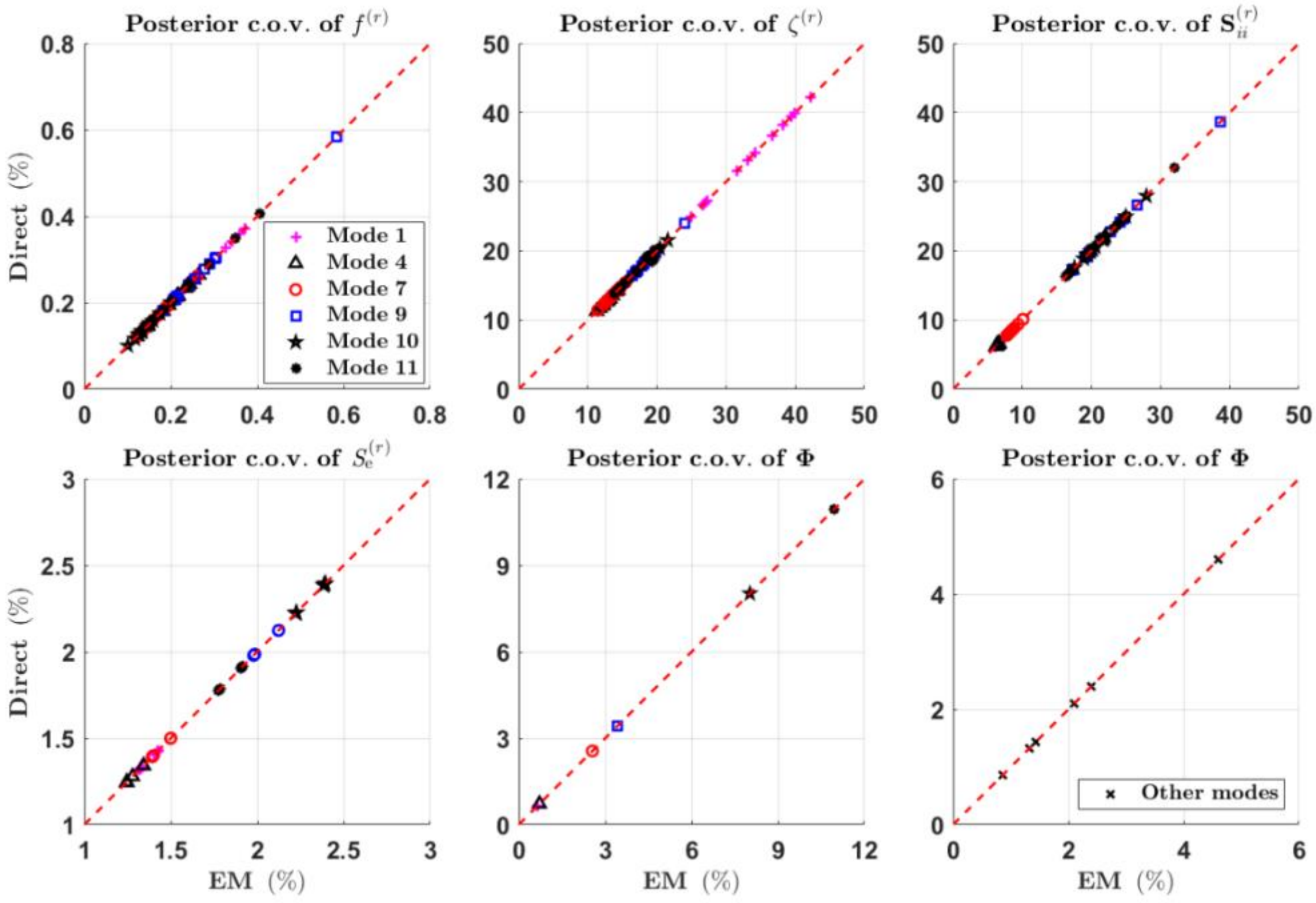

**Fig. 11.** Comparison of identification uncertainty, YR-ICC Building.

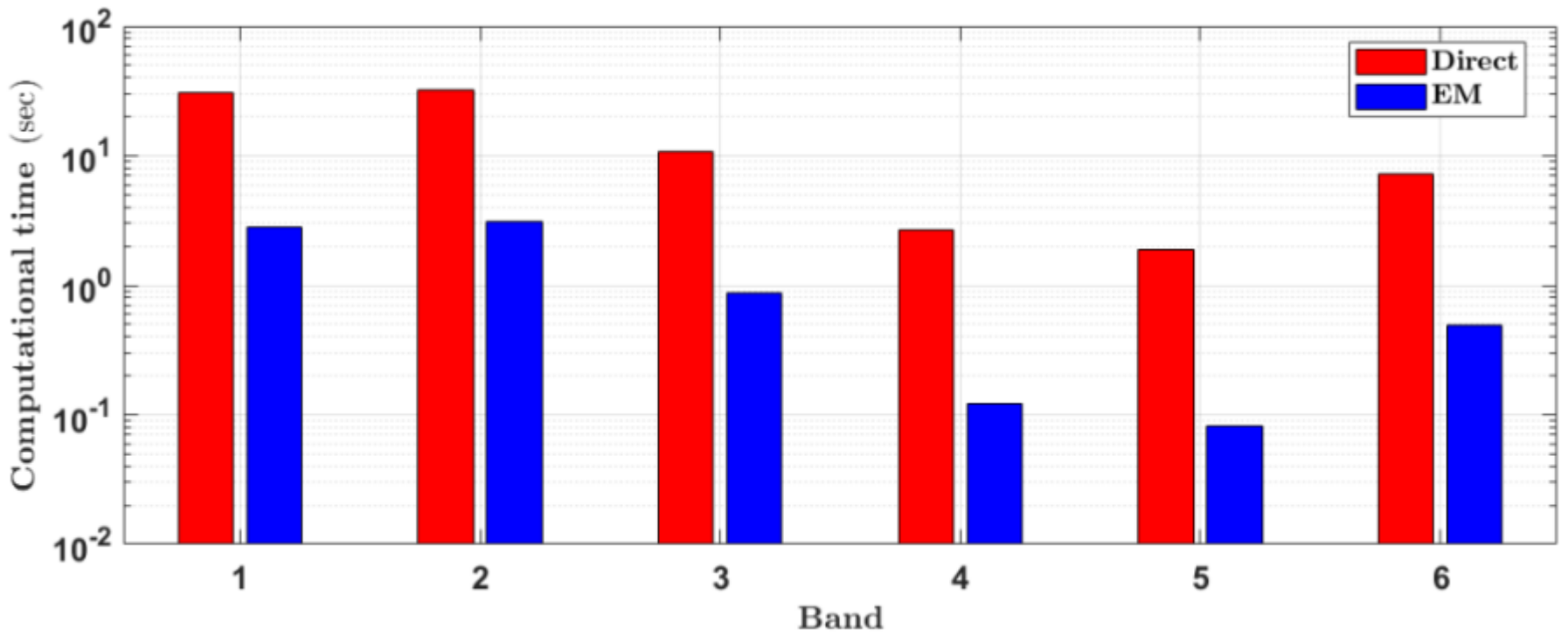

**Fig. 12.** Comparison of computational time, YR-ICC Building.

The sparsity patterns of the selection matrix and the Hessian matrix in this example are illustrated in **Fig. 13**. The first subfigure reveals that, although the selection matrix contains over 4800 elements, only 21 or 24 nonzero ones, which account for less than 0.5%. As a result, directly performing the direct multiplication outlined in Eqns. (19)~(21) would lead to significant memory inefficiency and increase computational burden. For the Hessian matrix shown in the next subfigure, we demonstrate the sparsity using three representative bands, containing one, two, and three modes, respectively. The sparseness increases as the matrix dimension (i.e., the number of parameters) grows, but the percentage of non-zero elements always remains around 13%. In the most challenging case, where three closely-spaced modes are covered, only 13.08% of the elements are nonzero. Given that this Hessian matrix is obtained by summing the counterparts over several hundred FFT points within the band, the sparse assembly strategy presented in Section 3.3 becomes essential for improving computational efficiency and conserving memory.

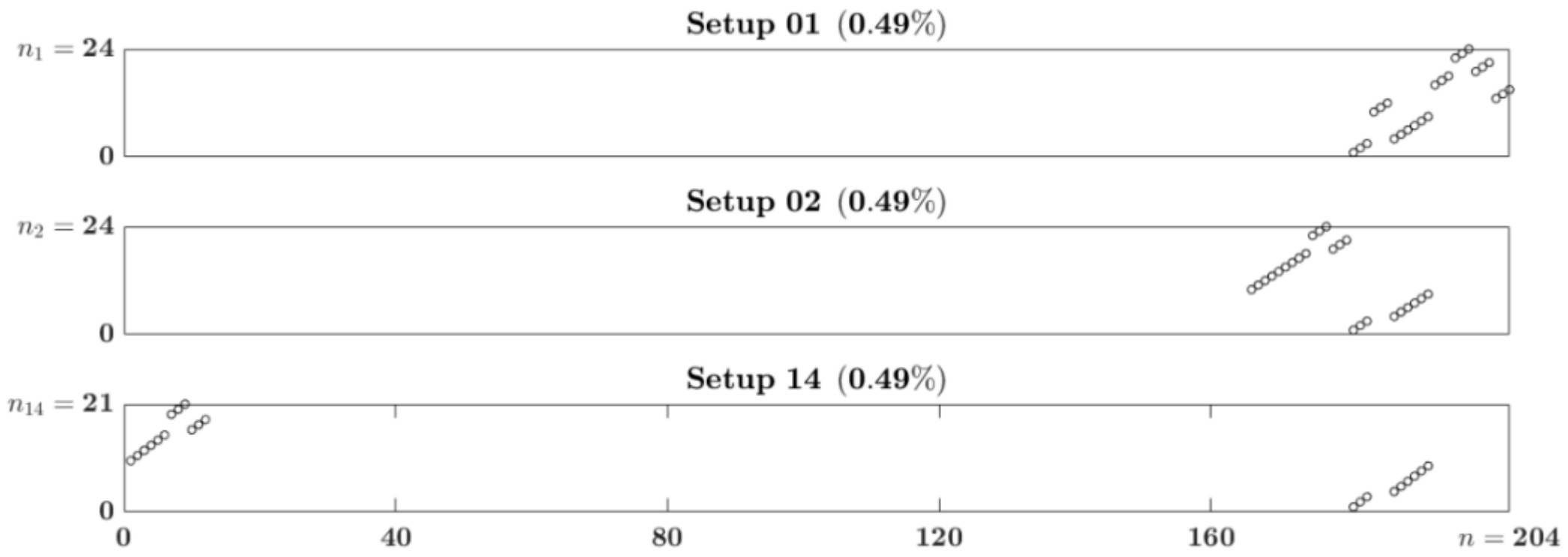

a) Nonzero elements in the selection matrix.

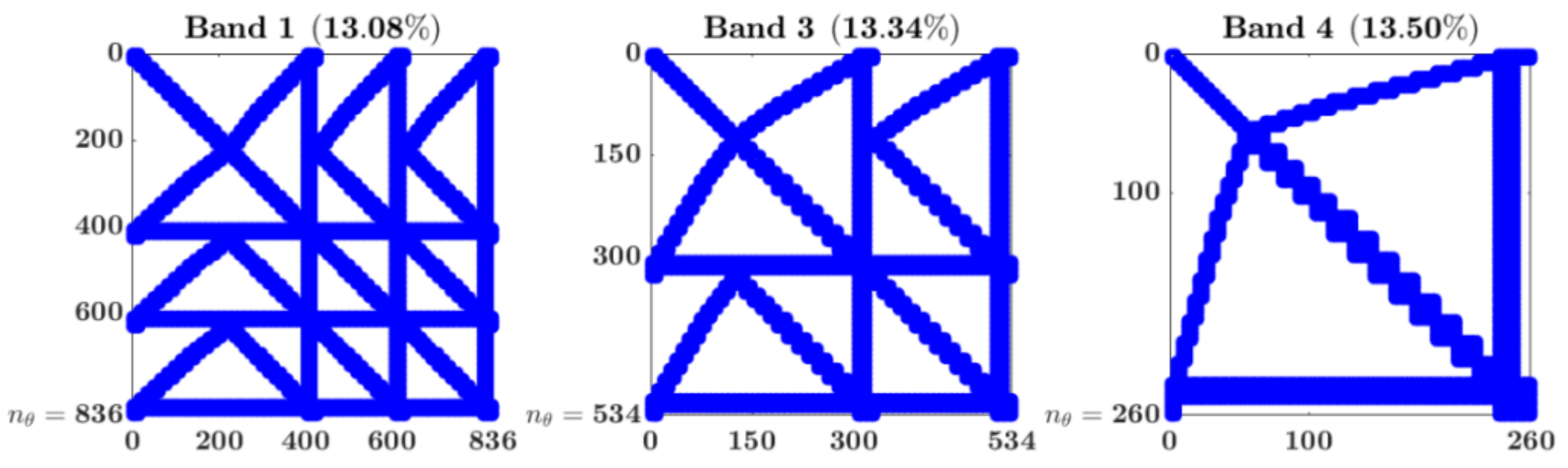

b) Nonzero elements in the Hessian matrix

**Fig. 13.** Sparse patterns visualization, YR-ICC Building. The number in the bracket denotes the percentage of the nonzero element.

# 6 Conclusions

An EM-aided approach for fast PCM computation has been developed for the Bayesian FFT method to quantify the identification uncertainty in multi-setup OMA. Leveraging the relationship between the NLLF and complete-data NLLF, the Hessian matrix is derived indirectly through Louis' identity, simplifying the computational process. A major advantage of this formulation lies in the maintained compatibility with the single-setup OMA expressions, enabling a straightforward extension to multi-setup OMA. Additionally, the proposed algorithm exploits the sparsity of the Hessian matrix, reducing memory usage and computational load. Such efficiency makes the approach highly effective for analyzing large-scale structures with closely spaced modes.

The accuracy and efficiency of the proposed algorithm have been verified using data from both a synthetic example and the YR-ICC building. It yields results identical to the existing approach, but with one-tenth of the computational time, even in challenging cases with tens of setups and closely-spaced modes. This performance demonstrates the potential for real-world modal analysis. Furthermore, integrating fast Hessian matrix computation into a Newton-based algorithm offers a promising avenue to further accelerate the Bayesian FFT method, advancing its real-time application in large-scale structural assessments.

# Declaration of competing interest

The authors declare that they have no known competing financial interests or personal relationships that could have appeared to influence the work reported in this paper.

# Declaration of generative AI and AI-assisted technologies in the writing process

During the preparation of this work the authors used ChatGPT in order to polish the language and improve the clarity of the text. After using this tool, the authors reviewed and edited the content as needed and take full responsibility for the content of the publication.

## Data availability

All data included in this study will be publicly available once it is accepted for publication.

## Acknowledges

This research is supported by the National Natural Science Foundation of China (U23A20662). The third author gratefully acknowledges the support of the UK Engineering and Physical Sciences Research Council (EPSRC) through the ROSEHIPS project (Grant EP/W005816/1). Any opinions, findings, conclusions, or recommendations expressed in this material are those of the authors and do not reflect the views of the funders.

## Appendix A. Matrix theory

This appendix provides the main results in matrix theory that are essential for deriving the formulas in this paper. For further details, please refer to Petersen and Pedersen [46].

1. Cyclic Property of Matrix Trace

The trace of a square matrix, denoted by $\mathrm{tr}(\cdot)$, is the sum of its diagonal elements. A fundamental property of the matrix trace is its cyclic invariance. For square matrices $\mathbf{A}$, $\mathbf{B}$, and $\mathbf{C}$ of appropriate dimensions, the cyclic property holds as

$$\mathrm{tr}(\mathbf{ABC}) = \mathrm{tr}(\mathbf{BCA}) = \mathrm{tr}(\mathbf{CAB}) \tag{A.1}$$

This property extends to any number of matrices. It is useful in simplifying expressions involving the trace of matrix products, as it allows for the rearrangement of matrices within the trace without changing the result.

2. Vectorization and Kronecker product

The vectorization operation denoted as $\mathrm{vec}(\cdot)$, is a linear transformation that converts a matrix into a column vector. This operation is useful in matrix calculus, where matrices are changed into vector forms to simplify manipulation and computation. Specifically, if $\mathbf{A} = [\boldsymbol{a}_1, \boldsymbol{a}_1, \cdots, \boldsymbol{a}_n] \in \mathbb{C}^{m\times n}$, the vectorization of $\mathbf{A}$ is defined as

$$\mathrm{vec}(\mathbf{A}) = \left[\boldsymbol{a}_1^{\mathrm{T}}, \boldsymbol{a}_2^{\mathrm{T}}, \cdots, \boldsymbol{a}_n^{\mathrm{T}}\right]^{\mathrm{T}} \in \mathbb{C}^{mn\times 1} \tag{A.2}$$

where $\boldsymbol{a}_j \in \mathbb{C}^{m\times 1}$ represents the $j$-th column of $\mathbf{A}$

The Kronecker product is a matrix operation that extends the outer product from vectors to matrices. Given matrices $\mathbf{A} \in \mathbb{C}^{m\times n}$ and $\mathbf{B} \in \mathbb{C}^{p\times q}$, the Kronecker product, denoted by $\mathbf{A}\otimes\mathbf{B}$, results in a larger matrix formed by multiplying each element of $\mathbf{A}$ by the entire matrix $\mathbf{B}$

$$\mathbf{A}\otimes\mathbf{B}=\begin{bmatrix} a_{11}\mathbf{B} & a_{12}\mathbf{B} & \cdots & a_{1n}\mathbf{B} \\ a_{21}\mathbf{B} & a_{22}\mathbf{B} & \cdots & a_{2n}\mathbf{B} \\ \vdots & \vdots & \ddots & \vdots \\ a_{m1}\mathbf{B} & a_{m2}\mathbf{B} & \cdots & a_{mn}\mathbf{B} \end{bmatrix}\in\mathbb{C}^{mp\times nq} \tag{A.3}$$

The manipulation imposed on the Kronecker product can be distributed to each individual term without changing the order, i.e.,

$$(\mathbf{A}\otimes\mathbf{B})^{\ddagger}=\mathbf{A}^{\ddagger}\otimes\mathbf{B}^{\ddagger} \tag{A.4}$$

where $[\cdot]^{\ddagger}$ can be transpose, conjugate, inverse, or their combinations.

These two operations are often used together to expand the dimensionality of matrices while preserving their structural properties. Introducing a new matrix $\mathbf{X}\in\mathbb{C}^{n\times p}$, a useful identity linking them is given by

$$\mathrm{vec}(\mathbf{AXB})=(\mathbf{B}^{\mathrm{T}}\otimes\mathbf{A})\mathrm{vec}(\mathbf{X}) \tag{A.5}$$

This identity becomes straightforward when $\mathbf{X}$ is the identity matrix, yielding

$$\mathrm{vec}(\mathbf{AB})=(\mathbf{B}^{\mathrm{T}}\otimes\mathbf{I}_m)\mathrm{vec}(\mathbf{A})=\left(\mathbf{I}_q\otimes\mathbf{A}\right)\mathrm{vec}(\mathbf{B}) \tag{A.6}$$

To vectorize the transpose of matrix $\mathbf{A}$, denoted as $\mathrm{vec}(\mathbf{A}^{\mathrm{T}})$, we can express it in terms of $\mathrm{vec}(\mathbf{A})$ using a commutation matrix $\mathbf{K}_{mn}$ [47], as follows

$$\mathrm{vec}(\mathbf{A}^{\mathrm{T}})=\mathbf{K}_{mn}\mathrm{vec}(\mathbf{A}) \tag{A.7}$$

where $\mathbf{K}_{mn}$ is essentially a permutation matrix determined only by the dimension of the matrix $\mathbf{A}$.

3. Hadamard Product

The Hadamard product, denoted as $\odot$, is an element-wise operation between two matrices of the same dimensions. If $\mathbf{C}=\left[c_{ij}\right]$ and $\mathbf{D}=\left[d_{ij}\right]$ are matrices in $\mathbb{C}^{m\times n}$, the Hadamard product of $\mathbf{C}$ and $\mathbf{D}$ is given by

$$\mathbf{C}\odot\mathbf{D}=\left[c_{ij}d_{ij}\right] \tag{A.8}$$

Specially, consider three square matrices $\mathbf{A}_1$, $\mathbf{A}_2$ and $\mathbf{A}_3$ in $\mathbb{C}^{m\times m}$ where $\mathbf{A}_2$ is diagonal. If the Hadamard product is taken with an identity matrix $\mathbf{I}_m$ in proper size, the order of multiplication does not affect the result, i.e.,

$$\mathbf{A}_1\mathbf{A}_2\mathbf{A}_3\odot\mathbf{I}_m=\mathbf{A}_2\mathbf{A}_1\mathbf{A}_3\odot\mathbf{I}_m=\mathbf{A}_1\mathbf{A}_3\mathbf{A}_2\odot\mathbf{I}_m \tag{A.9}$$

This invariance occurs because the Hadamard product with the identity matrix $\mathbf{I}_m$ extracts only the diagonal entries of the product $\mathbf{A}_1\mathbf{A}_2\mathbf{A}_3$. Therefore, the diagonal matrix only affects the diagonal elements when multiplied, and these elements are the same regardless of the order of multiplication. This property can be generalized for any sequence of square matrices $\mathbf{A}_1, \mathbf{A}_2, \cdots, \mathbf{A}_n$ where some matrices are diagonal. The order of these diagonal matrices can be arbitrarily rearranged without affecting the result when the Hadamard product is with the identity matrix. However, this property does not hold if the Hadamard product is with any matrix other than the identity matrix.

4. Matrix calculus

For any complex-valued variable $z = u + \mathbf{i}v$, the derivatives of a scalar function $h(z)$ with respect to (w. r. t.) $z$ and $z^*$ are defined as

$$\begin{aligned} \frac{\partial h(z)}{\partial z} &\triangleq \frac{1}{2}\left(\frac{\partial h(z)}{\partial u} - \mathbf{i}\frac{\partial h(z)}{\partial v}\right) \\ \frac{\partial h(z)}{\partial z^*} &\triangleq \frac{1}{2}\left(\frac{\partial h(z)}{\partial u} + \mathbf{i}\frac{\partial h(z)}{\partial v}\right) \end{aligned} \tag{A.10}$$

These partial derivatives are called Wirtinger derivatives, which treat a complex-valued variable and its conjugate as independent pairs.

Building on this, derivatives of scalar functions $h(\boldsymbol{z})$ and $h(\boldsymbol{Z})$ w. r. t. vector variable $\boldsymbol{z} \in \mathbb{C}^{n\times 1}$ and matrix variable $\boldsymbol{Z} \in \mathbb{C}^{m\times n}$ are defined as

$$\frac{\partial h(\boldsymbol{z})}{\partial \boldsymbol{z}} = \nabla_{\boldsymbol{z}} h(\boldsymbol{z}) \triangleq \begin{bmatrix} \frac{\partial h}{\partial z_1} & \frac{\partial h}{\partial z_2} & \cdots & \frac{\partial h}{\partial z_n} \end{bmatrix} \in \mathbb{C}^{1\times n} \tag{A.11}$$

$$\frac{\partial h(\boldsymbol{Z})}{\partial \boldsymbol{Z}} = \nabla_{\boldsymbol{Z}} h(\boldsymbol{Z}) \triangleq \frac{\partial h(\boldsymbol{Z})}{\partial \mathrm{vec}(\boldsymbol{Z})} = \begin{bmatrix} \frac{\partial h}{\partial z_{11}} & \frac{\partial h}{\partial z_{21}} & \cdots & \frac{\partial h}{\partial z_{m1}} & \cdots & \frac{\partial h}{\partial z_{mn}} \end{bmatrix} \in \mathbb{C}^{1\times mn} \tag{A.12}$$

Subsequently, we proceed to define the derivative of a vector function $\boldsymbol{h}(\boldsymbol{z}) \in \mathbb{C}^{m\times 1}$ w. r. t. $\boldsymbol{z}$

$$\frac{\partial \boldsymbol{h}(\boldsymbol{z})}{\partial \boldsymbol{z}} = \nabla_{\boldsymbol{z}} \boldsymbol{h}(\boldsymbol{z}) \triangleq \begin{bmatrix} \partial h_1/\partial x_1 & \partial h_1/\partial x_2 & \cdots & \partial h_1/\partial x_n \\ \partial h_2/\partial x_1 & \partial h_2/\partial x_2 & \cdots & \partial h_2/\partial x_n \\ \vdots & \vdots & \ddots & \vdots \\ \partial h_m/\partial x_1 & \partial h_m/\partial x_2 & \cdots & \partial h_m/\partial x_n \end{bmatrix} \in \mathbb{C}^{m\times n} \tag{A.13}$$

Finally, the derivative of a matrix function $\boldsymbol{h}(\boldsymbol{Z}) \in \mathbb{C}^{p\times q}$ w. r. t. $\boldsymbol{Z}$ is defined as

$$\frac{\partial \boldsymbol{h}(\boldsymbol{Z})}{\partial \boldsymbol{Z}} = \nabla_{\boldsymbol{Z}} \boldsymbol{h}(\boldsymbol{Z}) \triangleq \frac{\partial \mathrm{vec}[\boldsymbol{h}(\boldsymbol{Z})]}{\partial \mathrm{vec}(\boldsymbol{Z})} \in \mathbb{C}^{pq\times mn} \tag{A.14}$$

The chain rule in matrix calculus allows us to differentiate a function w. r. t. one variable by breaking it down into partial derivatives w. r. t. intermediate variables. It is useful when dealing with functions that involve matrix or vector inputs. Based on the definitions above, the chain rule for a function $\mathrm{g}(\mathbf{Y})$ where $\mathbf{Y} \in \mathbb{C}^{p\times q}$ is a function of $\mathbf{X} \in \mathbb{C}^{m\times n}$ is given by

$$\underbrace{\nabla_{\mathbf{X}}\,\mathrm{g}(\mathbf{Y})}_{1\times mn} = \underbrace{\nabla_{\mathbf{Y}}\,\mathrm{g}(\mathbf{Y})}_{1\times pq}\,\underbrace{\nabla_{\mathbf{X}}\mathbf{Y}(\mathbf{X})}_{pq\times mn} + \underbrace{\nabla_{\mathbf{Y}^*}\,\mathrm{g}(\mathbf{Y})}_{1\times pq}\,\underbrace{\nabla_{\mathbf{X}}\mathbf{Y}^*(\mathbf{X})}_{pq\times mn} \tag{A.15}$$

If $\mathbf{X} \in \mathbb{R}^{m\times n}$, the general expression of the second-order derivative $\nabla_{\mathbf{X}}^2\,\mathrm{g}(\mathbf{Y})$ using the chain rule can be written as

$$\underbrace{\nabla_{\mathbf{X}}^2\,\mathrm{g}(\mathbf{Y})}_{mn\times mn} = \underbrace{\nabla_{\mathbf{X}}^{\mathrm{T}}\mathbf{Y}(\mathbf{X})}_{mn\times pq}\,\underbrace{\nabla_{\mathbf{Y}}^2\,\mathrm{g}(\mathbf{Y})}_{pq\times pq}\,\underbrace{\nabla_{\mathbf{X}}\mathbf{Y}(\mathbf{X})}_{pq\times mn} + \underbrace{\nabla_{\mathbf{Y}}\,\mathrm{g}(\mathbf{Y}) \otimes \mathbf{I}_{mn}}_{mn\times mnpq}\,\underbrace{\nabla_{\mathbf{X}}^2\mathbf{Y}(\mathbf{X})}_{mnpq\times mn} \tag{A.16}$$

# Appendix B. Proof of zero-gradient of NLLF at MPV

This appendix gives an intuitive explanation of why the second term in Eqn. (15) can be ignored.

To begin with, we recall the MPV optimization procedures, especially how we account for the mode shape norm constraints. Given a starting point, an iterative algorithm like P-EM is applied to minimize the NLLF without considering the constraints first. Record the convergence result as a parameter set $\widehat{\boldsymbol{\theta}}' = \left[\widehat{\boldsymbol{x}}^{(1)}; \widehat{\boldsymbol{x}}^{(2)}; \cdots; \widehat{\boldsymbol{x}}^{(n_s)}; \mathrm{vec}\left(\widehat{\boldsymbol{\Phi}}'\right)\right]$, where $\boldsymbol{x}^{(r)} = \left[\boldsymbol{f}^{(r)}; \boldsymbol{\zeta}^{(r)}; \widehat{\mathbf{S}}^{(r)\prime}; \hat{S}_{\mathrm{e}}^{(r)}\right] \in \mathbb{R}^{n_{\theta_r}\times 1}$ are the independent variables in different setups. Note in this stage, the column norms of $\widehat{\boldsymbol{\Phi}}'$ need not be 1. Suppose the algorithm we employ decreases the NLLF adequately, yielding a zero-gradient at $\widehat{\boldsymbol{\theta}}'$, i.e., $\nabla L(\boldsymbol{\theta})|_{\boldsymbol{\theta}=\widehat{\boldsymbol{\theta}}'} = 0$. Then the renormalization of $\widehat{\boldsymbol{\Phi}}'$ and $\widehat{\mathbf{S}}^{(r)\prime}$ in Eqn. (13) does not change the NLLF value while satisfying the constraints. So the problem lies in whether the renormalized results can still generate a zero-gradient of the NLLF.

According to the Fisher's identity in Eqn. (25), the NLLF and the Q-function constructed in the EM algorithm share the same gradient. The analytical expression of the gradient of the Q-function has been shown in **Table 1**, which is related to the first and second moments of the latent variable. They are expressed by the convergence result $\widehat{\boldsymbol{\theta}}'$ as

$$\widehat{\boldsymbol{w}}_k^{(r)\prime} = \left[\widehat{\mathbf{P}}_k^{(r)\prime}\right]^{-1} \widehat{\boldsymbol{\Phi}}_r^{\prime\mathrm{T}}\widehat{\boldsymbol{\mathcal{F}}}_k^{(r)} \tag{B. 1}$$

$$\widehat{\mathbf{W}}_k^{(r)\prime} = \left[\widehat{\mathbf{P}}_k^{(r)\prime}\right]^{-1} \widehat{\boldsymbol{\Phi}}_r^{\prime\mathrm{T}}\widehat{\boldsymbol{\mathcal{F}}}_k^{(r)}\widehat{\boldsymbol{\mathcal{F}}}_k^{(r)\mathrm{H}}\widehat{\boldsymbol{\Phi}}_r^{\prime}\left[\widehat{\mathbf{P}}_k^{(r)\prime}\right]^{-1} + \hat{S}_{\mathrm{e}}^{(r)}\left[\widehat{\mathbf{P}}_k^{(r)\prime}\right]^{-1} \tag{B. 2}$$

where $\widehat{\mathbf{P}}_k^{(r)\prime} = \hat{S}_{\mathrm{e}}^{(r)}\left[\mathbf{h}_k^{(r)}\mathbf{S}^{(r)\prime}\mathbf{h}_k^{(r)*}\right]^{-1} + \widehat{\boldsymbol{\Phi}}_r^{\prime\mathrm{T}}\widehat{\boldsymbol{\Phi}}_r^{\prime}$; $\widehat{\boldsymbol{\Phi}}_r^{\prime} = \mathbf{C}_r\widehat{\boldsymbol{\Phi}}'$. Substituting the renormalization, i.e., Eqn. (13), yields $\widehat{\mathbf{P}}_k^{(r)} = \hat{S}_e^{(r)}\left[\mathbf{h}_k^{(r)}\mathbf{D}\mathbf{S}^{(r)\prime}\mathbf{D}\mathbf{h}_k^{(r)*}\right]^{-1} + \mathbf{D}^{-1}\widehat{\boldsymbol{\Phi}}_r^{\prime\mathrm{T}}\widehat{\boldsymbol{\Phi}}_r^{\prime}\mathbf{D}^{-1} = \mathbf{D}^{-1}\widehat{\mathbf{P}}_k^{(r)}\mathbf{D}^{-1}$. Therefore, the moments are updated as

$$\widehat{\boldsymbol{w}}_k^{(r)} = \mathbf{D}\left[\widehat{\mathbf{P}}_k^{(r)\prime}\right]^{-1}\mathbf{D}\mathbf{D}^{-1}\widehat{\boldsymbol{\Phi}}_r^{\prime\mathrm{T}}\widehat{\boldsymbol{\mathcal{F}}}_k^{(r)} = \mathbf{D}\widehat{\boldsymbol{w}}_k^{(r)\prime} \tag{B. 3}$$

$$\widehat{\mathbf{W}}_k^{(r)} = \mathbf{D}\widehat{\boldsymbol{w}}_k^{(r)\prime}\widehat{\boldsymbol{w}}_k^{(r)\prime\mathrm{T}}\mathbf{D} + \mathbf{D}\hat{S}_e^{(r)}\left[\widehat{\mathbf{P}}_k^{(r)\prime}\right]^{-1}\mathbf{D} = \mathbf{D}\widehat{\mathbf{W}}_k^{(r)\prime}\mathbf{D} \tag{B. 4}$$

Further substituting into **Table 1**, the first-order derivative w. r. t. $\boldsymbol{f}^{(r)}$ and $\boldsymbol{\zeta}^{(r)}$ updated as

$$\begin{aligned}\nabla_{\boldsymbol{f}^{(r)}}\hat{L}_k^{(r)} = 2\mathrm{Re}&\left[-\left(\frac{2\hat{\boldsymbol{f}}^{(r)\mathrm{T}}}{\mathrm{f}_k^{(r)2}} + \frac{2\mathbf{i}\hat{\boldsymbol{\zeta}}^{(r)\mathrm{T}}}{\mathrm{f}_k^{(r)}}\right)\left(\mathbf{D}\widehat{\mathbf{W}}_k^{(r)\prime}\mathbf{D}\left[\hat{\mathbf{h}}_k^{(r)*}\right]^{-1}\mathbf{D}^{-1}\left[\hat{\mathbf{S}}^{(r)}\right]^{-1}\mathbf{D}^{-1}\odot\mathbf{I}_m\right)\right] \\ &+\left(\frac{4\hat{\boldsymbol{f}}^{(r)\mathrm{T}}}{\mathrm{f}_k^{(r)2}} - \frac{4\hat{\boldsymbol{f}}^{(r)3\mathrm{T}}}{\mathrm{f}_k^{(r)4}} - \frac{8\hat{\boldsymbol{f}}^{(r)\mathrm{T}}\odot\hat{\boldsymbol{\zeta}}^{(r)2\mathrm{T}}}{\mathrm{f}_k^{(r)2}}\right)\widehat{\mathbf{D}}_k^{(r)} = \nabla_{\boldsymbol{f}^{(r)}}\hat{L}_k^{(r)\prime}\end{aligned} \tag{B. 5}$$

$$\begin{aligned}\nabla_{\boldsymbol{\zeta}^{(r)}}\hat{L}_k^{(r)} = 2\mathrm{Re}&\left[-\frac{2\mathbf{i}}{\mathrm{f}_k^{(r)}}\hat{\boldsymbol{f}}^{(r)\mathrm{T}}\left(\mathbf{D}\widehat{\mathbf{W}}_k^{(r)\prime}\mathbf{D}\left[\hat{\mathbf{h}}_k^{(r)*}\right]^{-1}\mathbf{D}^{-1}\left[\hat{\mathbf{S}}^{(r)}\right]^{-1}\mathbf{D}^{-1}\odot\mathbf{I}_m\right)\right] \\ &-\frac{8\hat{\boldsymbol{f}}^{(r)2\mathrm{T}}\odot\hat{\boldsymbol{\zeta}}^{\mathrm{T}}\widehat{\mathbf{D}}_k^{(r)}}{\mathrm{f}_k^{(r)2}} = \nabla_{\boldsymbol{\zeta}^{(r)}}\hat{L}_k^{(r)\prime}\end{aligned} \tag{B. 6}$$

where the invariance property in Eqn. (A.9) is used.

For $S_\mathrm{e}^{(r)}$, the relationship between the gradient before and after renormalization is given by

$$\begin{aligned}\nabla_{S_\mathrm{e}^{(r)}}\hat{L}_k^{(r)} &= 1/\left(n_r S_\mathrm{e}^{(r)}\right) - \widehat{\boldsymbol{\mathcal{F}}}_k^{(r)\mathrm{H}}\widehat{\boldsymbol{\mathcal{F}}}_k^{(r)}/S_\mathrm{e}^{(r)2} - \mathrm{tr}\left(\mathbf{D}\widehat{\mathbf{W}}_k^{(r)\prime}\mathbf{D}\mathbf{D}^{-1}\widehat{\boldsymbol{\Phi}}_r^{\prime\mathrm{T}}\widehat{\boldsymbol{\Phi}}_r^{\prime}\mathbf{D}^{-1}\right)/S_\mathrm{e}^{(r)2} \\ &+\left(\widehat{\boldsymbol{\mathcal{F}}}_k^{(r)\mathrm{H}}\widehat{\boldsymbol{\Phi}}_r^{\prime}\mathbf{D}^{-1}\mathbf{D}\widehat{\boldsymbol{w}}_k^{(r)\prime} + \widehat{\boldsymbol{w}}_k^{(r)\prime\mathrm{H}}\mathbf{D}\mathbf{D}^{-1}\widehat{\boldsymbol{\Phi}}_r^{\prime\mathrm{T}}\widehat{\boldsymbol{\mathcal{F}}}_k^{(r)}\right)/S_\mathrm{e}^{(r)2} = \nabla_{S_\mathrm{e}^{(r)}}\hat{Q}_k^{(r)\prime}\end{aligned} \tag{B. 7}$$

where the cyclic property in Eqn. (A.1) is employed.

Similarly, the derivatives w. r. t. $\mathbf{S}^{(r)}$ and $\boldsymbol{\Phi}_r$ are updated as

$$\begin{aligned}\nabla_{\mathbf{S}^{(r)}}\hat{L}_k^{(r)} &= \mathrm{vec}^\mathrm{T}\left[\mathbf{D}^{-1}\left(\left[\hat{\mathbf{S}}^{(r)\prime}\right]^{-\mathrm{T}} - \left[\hat{\mathbf{S}}^{(r)\prime}\right]^{-\mathrm{T}}\left[\hat{\mathbf{h}}_k^{(r)*}\right]^{-1}\widehat{\mathbf{W}}_k^{(r)\prime\mathrm{T}}\left[\hat{\mathbf{h}}_k^{(r)}\right]^{-1}\left[\hat{\mathbf{S}}^{(r)\prime}\right]^{-\mathrm{T}}\right)\mathbf{D}^{-1}\right] \\ &= \nabla_{\mathbf{S}^{(r)}}\hat{L}_k^{(r)\prime}\left(\mathbf{D}^{-1}\otimes\mathbf{D}^{-1}\right)\end{aligned} \tag{B. 8}$$

$$\begin{aligned}\nabla_{\boldsymbol{\Phi}_r}\hat{L}_k^{(r)} &= 1/S_e^{(r)}\,\mathrm{vec}^\mathrm{T}\left[\left(-\widehat{\boldsymbol{\mathcal{F}}}_k^{(r)*}\widehat{\boldsymbol{w}}_k^{(r)\prime\mathrm{H}} - \widehat{\boldsymbol{\mathcal{F}}}_k^{(r)}\widehat{\boldsymbol{w}}_k^{(r)\prime\mathrm{H}} + \widehat{\boldsymbol{\Phi}}_r^{\prime}\left(\widehat{\mathbf{W}}_k^{(r)\prime} + \widehat{\mathbf{W}}_k^{(r)\prime\mathrm{T}}\right)\right)\mathbf{D}\right] \\ &= \nabla_{\boldsymbol{\Phi}_r}\hat{L}_k^{(r)\prime}\left(\mathbf{D}\otimes\mathbf{I}_n\right)\end{aligned} \tag{B. 9}$$

where the Eqns. (A.5) and (A.6) are applied.

As a conclusion, the gradient at the MPV is the product of the gradient at $\widehat{\boldsymbol{\theta}}'$ and some constant matrices. Since the latter gradient is zero after the efficient optimization algorithm, the former one should also be zero.